\documentclass[10pt-default,11pt]{article}
\usepackage{amsmath}
\usepackage{amsfonts}
\usepackage{amssymb}
\usepackage{geometry}

\input{tcilatex}
\begin{document}

\bigskip

\bigskip

{\LARGE Searching for High Temperature Superconductivity:}

{\LARGE From Mendeleev to Seiberg-Witten via Madelung}

{\LARGE and Beyond}

{\LARGE \medskip }

$\bigskip $Arkady L.Kholodenko$^{\ast }\ \ \ \ \ \ \ \ \ \ \ \ \ \ \ \ \ \ \
\ \ \ \ \ \ \ \ \ \ \ \ \ \ \ \ $

\textit{375 H.L.Hunter Laboratories, Clemson University, Clemson, SC, USA}

$^{\ast }$E-mail: string@clemson.edu

\_\_\_\_\_\_\_\_\_\_\_\_\_\_\_\_\_\_\_\_\_\_\_\_\_\_\_\_\_\_\_\_\_\_\_\_\_\_%
\_\_\_\_\_\_\_\_\_\_\_\_\ \ \ \ \ \ \ \ \ \ \ \ \ \ \ \ \ \ \ \ \ \ \ \ \ \
\ \ \ \ \ 

Recently, a noticeable progress \ had been achieved in the area of high
temperature

superconductors. The maximum temperature T$_{c}$ of 250$^{0}$K$(-23^{0}$C$)$
for LaH$_{10}$ and

288$^{0}$K (+15 $^{0}$C) for CSH$_{8}$ were reported at the megabar
pressures. The highest

possible T$_{c}^{\prime }s$ were achieved by employing hydrides of chemical
elements. Empirically,

many of these are made of Madelung-exceptional atoms. Here the theoretical

background is provided explaining this observation. The, thus far empirical,
Madelung

rule is controlling Mendeleev's law of periodicity. Although the majority of
elements

do obey this rule, there are some exceptions. Thus, it is of interest to
derive it and its

exceptions theoretically in view of experimental findings. As a by product,
such a study

yields some plausible explanation of the role of Madelung-exceptional atoms
in the

design of high T$_{c}^{\prime }s$ superconductors. Thus far the atoms
obeying the Madelung rule and

its exceptions were studied with help of the relativistic Hartree-Fock
calculations.

In this work\ we \ reobtain both the rule and the exceptions analytically.
The newly

developed methods are expected to be of value in quantum \ many-body theory
and, in

particular, in the theory of high T$_{c}$ superconductivity. Ultimately, new
methods involve

some uses of the Seiberg-Witten (S-W) theory known as the extended
Ginzburg-Landau

theory of superconductivity. \ Using results of the S-W theory the
difference between the

Madelung-regular and Madelung-exceptional atoms is explained in terms of the

topological transition. Extension of this, single atom, result to solids of
respective

elements is also discussed

\_\_\_\_\_\_\_\_\_\_\_\_\_\_\_\_\_\_\_\_\_\_\_\_\_\_\_\_\_\_\_\_\_\_\_\_\_\_%
\_\_\_\_\_\_\_

Subject index \ \bigskip \bigskip \bigskip \bigskip A 13, A 60, A 63, E0

\newpage

\textbf{1. Introduction} \bigskip

\textbf{1.1. Not too widely known facts about the atomic
superconductivity\bigskip }

\ \ The most essential feature of superconductivity is the Meissner
effect-the expulsion

of superconductor from the applied static magnetic field. Such feature makes
all

superconductors diamagnetic. With respect to the external magnetic field all
atoms

exhibit magnetic properties as well. It is not immediately clear though will
the magnetic

properties of individual atoms survive if they form the bulk solid phase. \
The hydrogen-the

simplest of chemical elements- already exhibits a variety of puzzling
properties. Although

it is studied in every course on quantum mechanics, standard treatments
imply that the

atomic hydrogen should be both paramagnetic (strongly) and diamagnetic
(weakly).

In reality, it is diamagnetic only. That is, already\ the atomic hydrogen is
a superconductor!

If this is so, will this property survive for the solid hydrogen?
Surprisingly, there is no

mathematically rigorous answer to this question.

\ The standard methods of quantum mechanics indicate that the paramagnetism
of

hydrogen should exist. And, indeed, all other hydrogen-like atoms, e.g. Li,
Na, K, Rb,

Cs are indeed paramagnetic as experiment and elementary calculation
demonstrate.

Furthermore, the surprising atomic diamagnetism of hydrogen is followed by
the much

anticipated diamagnetism of He and Be. These observations are misleading,
though.

Indeed, all noble gases are diamagnetic but so is Be but they are
practically not

superconducting in their bulk and Be under normal ambient pressures is very
bad

superconductor. Thus, bulk superconductivity implies diamagnetism but \ the
opposite

is not true. The diamagnetism of atomic hydrogen is a subject matter of
hundreds of

publications. It can be explained group-theoretically$[1]$, using the theory
of quantum

chaos $[2],$ or perturbation \ theory of superintegrable systems\footnote{%
E.g. read the definition of superintegrability in [3]. Study of the
sophisticated perturbational
\par
\ \ theory of\ superintegrable systems is also mentioned in [3].}.

\ \ \ The diamagnetism of atomic H and paramagnetism of Li, Na, K, Rb, Cs
seemingly

affects their phase diagrams. Li, Na, K, Rb, Cs readily become metals under
normal

pressures while for H this is not possible. In 1935 Eugen P.Wigner and \
Hillard B.

Huntigton [$4$] predicted that only at the pressures above 25 GPa hydrogen
would

become an alkali metal-like solid. Some (but not all!) of alkali metals
under similar

pressures will become superconducting [$5$]. In 1968 Ashcroft and,
independently,

Ginzburg, in 1969, predicted that the metalized hydrogen is an ideal
candidate for

exhibiting the high T$_{c}$ superconductivity [$5$]. Although up to this
moment the

superconducting hydrogen is not found yet reliably, many theoretical
predictions

do exist indicating that the superconducting hydrogen might exist at
pressures above

450-500 GPa, at temperatures at and above the room temperature. Its
superconductivity

is believed to be described by the well studied Bardeen-Cooper-Schriffer
(BCS) theory

and/or its Migdal-Eliashberg modification. The question immediately emerged:
Is there

a way to reduce the pressures for the hydrogen while retaining
superconductivity? The

important step in this direction was made by Gilman, in 1971 [$6$]. He
suggested to use

hydrides XH$_{n}$, where X is an atom other than hydrogen and $n$ is the
number of

hydrogen atoms attached to it, perhaps under high pressure\footnote{%
In practice $n$ may not be an integer}. Gilman's idea to use

hydrides instead of hydrogen was named \textbf{chemical precompression.}

\ \textbf{\ \ }The idea seemed very attractive because of the following
logic behind it. 1) Take an

element (metal or not), 2) apply a pressure to it till it becomes a solid,
3) at the stages

1) and/or 2) try to saturate this solid with the atomic hydrogen. Such a
saturation will

eventually create a hydrogen sublattice\footnote{%
There could be more than one sublattice} inside the host lattice. The
sublattice

will force hydrogen to act like a solid and, hopefully, this solid will be
superconducting

under mild readjustment of external parameters.

\ \ \ From the description of precompression several questions emerge. 1)
How to make an

alloy with prescribed number $n$ of hydrogen atoms? 2) Is there any
relationship between

$n$, T$_{c}$ and the pressures? 3) How stable are hydrogen sublattices? \
Since these topics

were discussed in [$5$] this spares us from the extended discussion. At the
same time, since

answering to these questions brings us directly to the subject matters of
this paper we

present some additional comments in the next subsection.\medskip

\textbf{1.2. \ Reversible hydrides. Peculiar interplay between the atomic
and }

\ \ \ \ \ \ \ \ \textbf{bulk\ superconductivities studied with Bogoliubov's
method of }

\ \ \ \ \ \ \ \ \textbf{quasiaverages}

\textbf{\ \ \ }

\ \ From the previous subsection we learned that hydrogen-the lightest among
atomic

superconductors- is expected to yield under very high pressures the highest
possible T$_{c}.$

Even if this result is achieved in laboratory, extremely high pressures make

such a project of academic interest only. We do not discuss here situations

inside of the stellar or planetary cores leading to the emergence of the
permanent

magnetic field around these objects. Thus, focusing attention on hydrides
makes more

sense. But the problem

\#1) from previous subsection still remains. And, hence, some requirements
should be

applied to make correct selection of the atom X.

\ According to [$5$] (page 45) "in order to obtain the effective
metallization of the hydrogen

sublattice, it would be more convenient to start from the existing
hydrogen-rich molecule

since in this case the hydrogen does not have to be incorporated \ into a
host metal lattice."

No examples of such hydrated molecules are given in [$5$]\footnote{%
However, based on the results of this paper hydrated molecules made of the
\par
Madelung-exceptional and hydrogen atoms are permissible theoretically.}. At
the same time,

in [$5$] the comments on \# 2) are as follows (page 5): "The possibility of
predicting T$_{c}$

from the first principles played crucial role in the second hydride
revolution (dawn of

2000s) as well as the development of computational tools to predict crystal
structures

and phase diagrams of materials under given thermodynamic conditions."
Nevertheless,

on page 47 of [$5$] we found the following clarification: "rare-earth and
actinides are

theoretically challenging to describe, and one can quickly obtain wrong
results..." But

exactly these elements are the Madelung -exceptional! They are also yielding
hydrides

with the highest T$_{c}^{^{\prime }}s$. This quotation from [$5$] provides
us with the first compelling reason

to study further the Madelung-exceptional elements. Furthermore, the
intuitive belief

that the larger $n$ is the higher \ should be the T$_{c}^{^{\prime }}s$ is
not always working for the following

reason. On page 20 of [$5$] we find: " systems containing light mass atoms,
like hydrides

under pressure, exhibit intrinsically large vibrational displacements and
hence show-

case a variety of effects due to strong anharmonicity..." Thus, if the high T%
$_{c}^{\prime }s$ cannot be

achieved just by increasing $n$, the focus of attention shifts to the
Madelung-exceptional

elements for the following reason. This reason is historical. It \textbf{not
at all} motivated by

the Madelung-exceptionality of elements.

\ \ \ In 1866 Thomas Graham discovered that at 1 atm \ metallic palladium
can absorb \ 

hydrogen in large amounts: 0.58 H atoms per one Pd atom. \ Since that time,
for more

than 150 years the Pd-H$_{2}$ system remains as the benchmark model for
studying the metal

hydrides, beginning with PdH$_{x}$[$7$]. For this historical reason, study
of superconducting

hydrides had began with hydrides of Pd and Th. Incidentally, although both
are the

Madelung-exceptional, nowhere in the literature this fact is
mentioned/emphasized.

The motivation thus far came from another observations. Under normal
atmospheric

pressure Pd is not a superconductor and Th is exceptionally bad
superconductor. It has

T$_{c}$=1.374 $^{0}$K [$8$]. At the same time, the hydride PdH$_{x}$ (x$%
\simeq 0.7)$ is a

superconductor, with T$_{c}\simeq 9^{0}$K and for ThH$_{3.84}$ the T$_{c}$
was found as 8.35$^{0}$K [$9$]. These

results demonstrate that use of hydrides is step in the right direction. The
problem,

nevertheless, remains. Yes, the PdH$_{x}$ and ThH$_{3.84}$ had noticeably
improved their T$_{c}^{\prime }s$

upon making hydrides and, for the PdH$_{x}$ the T$_{c}$ is increasing
linearly with the linear

increase in fraction of H in PdH$_{x}$ ([$5$], page 49). But, is it possible
to regulate the

the amount of H in hydrides of other atoms to the extent it was done in Pd
and Th ?

Yes, computational advancements formally allow to make predictions of T$%
_{c}^{\prime }s$ for the

assigned pressure, but to what extent these hydrides can be recreated in
real life? And,

as it was mentioned above, in this subsection, for the Madelung-exceptional
atoms

computations are not reliable but the obtained T$_{c}^{\prime }s$ are among
the highest. Because of

this we shall focus our attention on Madelung-exceptional atoms. In doing
so, we shall

cite extensively the content of Ref.s [$10$],[$11$].

\ \ \ According to [$10$], the molecular dissociation of H$_{2}$ is the
first step toward formation of

hydrides by absorption. Other than Pd, most metals require energy input in
order to

overcome the activation barrier. This is achieved by application of high
hydrogen

pressures and elevated temperatures. On Pd surfaces, the dissociative
absorption of

H$_{2}$ molecules \textbf{occurs with little or no activation energy barrier}%
! This fact causes

\textbf{the absorption to be reversible} and, therefore, following [$11$]
all hydrides for which

the absorption is reversible \ are called \textbf{reversible hydrides}.
Empirically, it is known

[$11$], page 31, that all the reversible hydrides \ working around \ ambient
temperature

and atmospheric pressure consist of transitional and rare earth metals, that
is almost

all of them are Madelung-exceptional! In particular, very good reversible
hydrides

are made of Pt and Ru [$11$]. What is the physics behind this phenomenon?
The hint

is given by the keyword: \textsl{reversible hydrides}.

\ \ To move forward, we need to have some model of the metal. This will
enable us to

describe the absorption-desorption process within limits of this model. The
simplest

model of metal is some weakly or strongly interacting electron gas on some
jellium-

like neutralizing background. It happens, that the description of processes
of

absorption-desorption based on such model [$12$] make good sense.
Irrespective to

ramifications of a given model, all models are subject to some constraints
of general

nature. These are associated with symmetry. Conservation of energy, momentum,

spin, etc. all are consequences of symmetry. More delicate are the
conservation laws

associated with, for example, conservation of particles. These are
associated with

(global) gauge invariance symmetry. Also, it matters wether particles are
bosons or

fermions. Photons, phonons, plasmons are bosons and they are massless. In

superconductivity two fermions forming a Cooper pair become one boson. But
this

BCS boson is massive. Bogoliubov developed a very general concept of

\textsl{quasiaverages }[$13$],[$14$] in connection with his seminal works on
superconductivity.

It is essential to emphasize that this concept is far more general than just
the

superconductivity problematics [$14$]. The concept of quasiaverages is useful

whenever there is some change in symmetry. In the present case, we are
dealing with

the fermion system whose number of particles is not conserved\footnote{%
E.g. hydride molecule in the simplest case.}.This is

indicative of spontaneous breakage of global U(1) symmetry associated with

electromagnetism. Within the framework of superconductivity, details are
provided

in section 6 and Appendix E.3. \ 

The case of reversible hydrides falls into this category. Indeed, the
absorption

process begins with H$_{2}$ breaking into 2 H's each having proton and
electron. Both are

fermions. When the pair of H's enters the bulk solid, it donates 2 electrons
(fermions) to

the interacting electron gas and 2 protons (fermions) to the jelly. Since
the process is

reversible, the 2 H's can emerge back at the surface. Since the bulk system
is not particle-

conserving such reversible process requires no energy for it to happen. Such
picture

is missing thus far one very important ingredient. Just described
description of

absorption-desorption is \textbf{not at all} valid for \textbf{all} solids
(metallic or not)! It is only valid

for solids made of the hydrides of Madelung-exceptional atoms since \textbf{%
already at }

\textbf{the atomic level these atoms ( and only these!) are superconducting }%
as

explained in the Appendices E.4 and E.5. In the previous subsection we
argued that

the H atom is also superconducting since it is diamagnetic. Since such
Madelung-

exceptional solids are reversible hydrides, this makes these solids \
nontraditional \ 

superconductors in a way just described. This makes sense because Pd is not a

superconductor without H's. Presence of H's in whatever amounts makes it

superconducting in conventional sense. The same is true for other Madelung-

exceptional elements. Just presented results allow us to formulate the
content

of the rest of this work.\medskip

\textbf{1.3. Organization of the rest of the paper\medskip }

In section 2 we present basic facts about the atomic physics allowing us to
introduce

definitions of the Madelung-regular and Madelung -exceptional atoms. In
section 3 we

initiate \ an explanation of what makes the Madelung-exceptional atoms
exceptional. We

argue, that: a) the Madelung-exceptionality is relativistic phenomenon, b)
application

of relativistic methods known in physics literature, makes all atoms
Madelung-exceptional.

This creates the first fundamental problem: how to disentangle the Madelung
-exceptional

atoms from the Madelung -regular? \ In section 3 we explicitly explain what
features make

the atoms Madelung-regular. In reading sections 2 and 3 our readers are
instructed

to read the appendices- A, B and C. The reading of these appendices is not
optional. \ 

Section 4 is \ meant to prepare our readers for new information. For this
purpose we had

converted results of section 3 into equivalent geometrical/topological form
allowing us

to account for effects of covariance, gauge invariance (local and global),
etc. This

conversion was influenced by the work of Schr\"{o}dinger on Dirac electron
in the

gravitational field. In doing so we used the original work by Schr\"{o}%
dinger [$30$],

written in German, as well as its English translation [$32$]. Results of
section 4 allow

us to bring into play the results of Seiberg-Witten (S-W) theory in section
5. Although

this theory, according to its author, E.Witten [$50$], is just a
sophisticated extension

of the Ginzburg-Landau theory of superconductivity, to our knowledge, there
were

no precedents, till this paper, to demonstrate explicitly (using physical
terminology),

the connection of the S-W formalism with that for superconductivity. In this
work it

is done with the purpose of demonstrating that mathematically the transition
from the

Madelung-regular to Madelung-exceptional atomic behavior is of topological

nature. The case of Madelung-regular atoms requires for its description the

concept of $spin$ manifold while the Madelung-exceptional atoms can "live"
only

on $spin^{c}$ manifolds. Since the $spin^{c}$ and $spin$ manifolds are
topologically different,

the transition from the Madelung-regular to Madelung-exceptional atoms is

topological in nature. The definition of $spin^{c}$ manifolds in
mathematical literature

[$70$] is devoid of any traces of physics. Being motivated by the physics of
reversible

hydrides, discussed in subsection 1.2., we found physical interpretation of $%
spin^{c}$

manifolds in terms of known concepts of BCS superconductivity, e.g. using the

Bogoliubov-De Gennes equations. Details are provided in the Appendices D and
E.

Finally, in the Appendix F we still further simplified the concept of $%
spin^{c}$

manifolds and tested this simplified definition using known examples of

electron filling patters\ for the Madelung-exceptional and regular atoms.
The S-W

formalism also allowed us to demonstrate, in the Appendix C, that the number
of

Madelung-exceptional elements is always finite and always holds only for
heavier

atoms, where the relativistic effects are non negligible.\medskip\ 

In section 6 \ we discuss two problems. These are: a) the problem of
extension

of just obtained single atom results to solids of macroscopic sizes; b) \
provided that

the problem a) is solved, will these solids remain \ superconductive?

Section 7 is devoted to the summary and discussion.\medskip \medskip

\textbf{2.} \textbf{Some facts about the periodic system of elements. }

\ \ \ \ \textbf{Madelung-regular vs Madelung-exceptional atoms\medskip
\medskip }

\ \ Although quantum mechanical description of multielectron atoms and
molecules is

considered to be a well developed domain of research, recently published
book $[15]$

indicates that there are many topics to be addressed still. The quantum
mechanical

description of multielectron atom (with atomic number $Z$ and infinitely
heavy nucleus)

begins with writing down the stationary Schr\"{o}dinger equation 
\begin{equation}
\hat{H}\Psi (\mathbf{r}_{1},\mathbf{r}_{2},...,\mathbf{r}_{Z})=E\Psi (%
\mathbf{r}_{1},\mathbf{r}_{2},...,\mathbf{r}_{Z})  \tag{1}
\end{equation}

with the Hamiltonian%
\begin{equation}
\hat{H}=-\dsum\limits_{i=1}^{Z}\frac{\hslash ^{2}}{2m}\nabla
_{i}^{2}-\dsum\limits_{i=1}^{Z}\frac{Ze^{2}}{r_{i}}+\frac{1}{2}\dsum\limits 
_{\substack{ i,j=1  \\ i\neq j}}^{Z}\frac{e^{2}}{r_{ij}}.  \tag{2}
\end{equation}

\ Bohr's \textsl{Aufbauprinzip} postulates that the atom with atomic number $%
Z$ is made of $Z$

electrons added in succession to the bare atomic nucleus. At the initial
stages of

this process the electrons are assumed to occupy the one-electron levels of
the lowest

energy. Mathematically, this process is described in terms of the one
electron

eigenvalue problem 
\begin{equation}
\hat{H}_{i}\psi _{\square _{i}}(\mathbf{r}_{i})=[-\frac{\hslash ^{2}}{2m}%
\nabla _{i}^{2}+V_{eff}(\mathbf{r}_{i})]\psi _{\square _{i}}(\mathbf{r}%
_{i})=\varepsilon _{nl}(i)\psi _{\square _{i}}(\mathbf{r}_{i}),i=1\div Z, 
\tag{3}
\end{equation}

where $V_{eff}(\mathbf{r}_{i})$ is made of the combined nuclear potential - $%
\frac{Ze^{2}}{r_{i}}$ and the centrally symmetric

Hartree-Fock-type potential $\mathcal{F}$(\textbf{r}$_{i})$ for the i-th
electron coming from the presence of the

rest of atomic electrons. The fact that $\mathcal{F}$(\textbf{r}$_{i})$ is
indeed \textbf{centrally-symmetric} was

demonstrated in the book by Bethe and Jackiw $[16]$.\textbf{\ It is
fundamentally important}

for our calculations. The symbol $\square _{i}$ indicates the i-th entry
into the set\ made out of

hydrogen-like quantum numbers for individual \ electrons. Based on this, the
concept

of\textsl{\ an orbital} is associated with the major quantum number $n$
having its origin in studies

of hydrogen atom. In quantum many-body system described by Eq.(3) it makes
more

sense however to associate the concept of an \textsl{orbital }with the
description of somehow

labeled, say, by interaction with photon (when studied spectroscopically),
the i-th electron

moving in the centrally symmetric potential $V_{eff}(\mathbf{r}_{i}).$ 
\textsl{The quantum motion in such a}

\textsl{potential should cause the hydrogen quantum numbers} $n,l,m$ \textsl{%
and} $m_{s}$ \textsl{to change into}

\textsl{hydrogen-like\footnote{%
E.g.read section 3.3.}} since the hydrogen atom eigenvalue problem is now
being replaced by

the eigenvalue problem for the labeled i-th electron in the centrally
symmetric potential

$V_{eff}(\mathbf{r}_{i})$ which is different from the Coulombic. The actual
implementation of this

observation is presented in this work from the new standpoint. It is known
that the

number of electrons allowed to sit on such redefined orbital is determined
by the \textsl{Pauli}

\textsl{exclusion principle. }With increasing $Z$ the electrons are expected
to occupy the

successive orbitals according to Bohr's Aufbau scheme until the \textsl{%
final ground state }

\textsl{electron configuration }is reached. This is achieved by using the
assumption made

by Bohr that the atom with $Z$ electrons is made out of atom with $Z-1$
electrons by

a) changing the nuclear charge by +1 and, by simultaneously adding one
additional

electron. In such imaginary process it is assumed that \ the quantum numbers
of electrons

in the $Z-1$ atom remain unchanged $[17]$.

\ \ \ The problem with \textsl{Aufbauprinzip, }just described, lies in the
assumption that the

guiding principle in designing the final ground state electron configuration
is made

out of two components:

a) knowledge of hydrogen-like wave functions supplying (labeled by) the
quantum

boxes/numbers $\square _{i}$ and,

b) the Pauli principle mathematically restated in the form of fully
antisymmetric

wavefunction $\Psi (\mathbf{r}_{1},\mathbf{r}_{2},...,\mathbf{r}_{Z}).$
Although mathematically it is just an exterior

differential form, the existing treatments do not use the Hodge-De Rham
theory of

differential forms for description of Pauli principle. In this work this is
to be corrected. \ 

Should the a) and b) requirements be sufficient, then the familiar
hydrogen-like

quantum numbers $n,l,m$ and $m_{s}$ for the hydrogen would make the filling
of electronic

levels \ to proceed according to the Fock $n$-rule.\medskip

\textbf{Fock n-rule: }\textsl{With increasing Z the nl orbitals are filled
in order of }

\textsl{increasing n.\medskip }

This rule leads to problems already for the lithium [$15$], page 330. As
result, the n-rule

was replaced by the ($n,l$) rule.\medskip

\textbf{The hydrogenic (}\textit{n},\textit{l})\textbf{\ rule: }\textsl{With
increasing Z, the orbitals are }

\textsl{filled in order of \ increasing n while for a fixed n the orbitals }

\textsl{are filled in order of increasing \textit{l}.\medskip }

After $Z=18$ the ($n,l$) rule breaks down as well. Therefore, it was
subsequently

replaced by the ($n+l,n$) rule of Madelung\medskip .

\textbf{The Madelung (}\textit{n+l},\textit{n}\textbf{) rule: }\textsl{With
increasing Z, the orbitals are }

\textsl{filled in order of \ increasing n+}$\QTR{sl}{l}=N.$ \textsl{For the
fixed }$N$\textsl{, the orbitals}

\textsl{are filled in order of increasing n.\medskip }

This rule was included by Madelung in his 1936 book$[17]$ in the form of an
Appendix 11

describing the filling of periodic table. In the same Appendix 11 Madelung
confesses

that: a) the filling rule is strictly empirical and, b) as such, it does
possesses some

exceptions. The Madelung rule and its exceptions require theoretical
explanations.

\ \ \ \ By organizing the elements in periods of constant $n+l$ and, in
groups of constant

$l,m_{l}$ and $m_{s}$, the period doubling emerges naturally and leads to
the sequence of periods:

2-2-8-8-18-18-32-32. Using the apparatus of dynamical group theory\ in $[15]$
\ the period

doubling was recreated. Application of the group-theoretic methods to
periodic system \ 

was done repeatedly in the past. To our knowledge, the most notable

results are presented in Chapter 6 of the book by Englefield[$18$]. \ Much
later, the results

of Chapter 6 were independently reobtained in$[15]$. Should the Madelung
rule be without

exceptions, just mentioned results would be sufficient. However, the
existing exceptions

for some transition metals, lanthanides and actinides indicate that uses of
the dynamical

group theory methods alone are not sufficient. As result, in this work we
describe the

alternative methods \ enabling us to explain the Madelung rule and its
exceptions. \medskip \medskip

\ The problem of finding the theoretical explanation of the Madelung rule
had attracted

the attention of Demkov and Ostrovsky (D-O)[$19]$. They used methods, other
than

group-theoretic, enabling them to guess $V_{eff}(\mathbf{r}_{i}$) correctly.
This had been achieved by

taking into account implications of the Bertrand theorem of classical
mechanics [$20$].

It imposes apparently insurmountable restrictions on the selection of $%
V_{eff}(\mathbf{r}_{i}):$ for

spherically symmetric potentials only the Coulombic -$\frac{Ze^{2}}{r_{i}}$
and the harmonic oscillator

$kr^{2}$ potentials allow dynamically closed orbits. Theoretical treatment
of multielectron

atoms before D-O works was confined either to study of spectra of
classically and quantum

mechanically chaotic systems or to uses of variational (relativistic or not)
Hartree-Fock

spectral calculations. Beginning with the motion of electrons in helium
atom, the classical

(and, hence, the semiclassical!) dynamics of \ electrons in multielectron
atoms is believed

to be chaotic. The seminal book by Gutzwiller [$21$] is an excellent \
introduction to this

topic. Already Bethe and Jackiw$[16]$ noticed that the Hartree-Fock $V_{eff}(%
\mathbf{r}_{i})$ is centrally

symmetric. This brings into question the issue of description of description
of

multielectron atom at the semiclassical level. D-O found seemingly
innovative approach

to the spectral problem. They applied the optical-mechanical analogy in
which the

Maxwell fish-eye potential was used instead of the Coulombic potential for
hydrogen

atom. D-O believed that such a replacement might help them to cope with the

multielectron effects while \ keeping an agreement with the Bertrand
theorem. To do so,

they: a) replaced the Coulomb potential by the fish-eye potential and b)
used the

conformal transformations applied to the fish-eye potential aimed at
conformally

deforming this potential in such a way that it will correctly represent the
multielectron

effects. At the level of classical mechanics D-O demonstrated the
equivalence (for the

hydrogen atom) between the Hamilton-Jacobi equations employing \ the Maxwell

fish -eye and \ the Coulombic potentials. In the Appendix B we reproduce
needed

details and comment on some flaws in D-O reasonings. At the quantum level

D-O believed that "The Maxwell's fish-eye problem is \textbf{closely related
to} the

Coulomb problem." \ Being aware of the book by Luneburg [$22$] D-O
nevertheless

underestimated the nature of the connection between the Coulombic and optical

(fish-eye) problems. The assumption of only "close relationship" caused D-O
to

replace Eq.(3) by%
\begin{equation}
\lbrack -\frac{\hslash ^{2}}{2m}\nabla _{i}^{2}+V_{eff}(\mathbf{r}_{i})]\psi
(\mathbf{r}_{i})=0.  \tag{4}
\end{equation}

Eq.(4) is looking differently from Eq.(3). Eq.(3) is an eigenvalue spectral
problem while

Eq.(4) is the Sturmian problem. That is to say, for the Sturmian-type
problem to be well

defined, the parameters entering into $V_{eff}(\mathbf{r}_{i})$ must be
quantized. Such quantization

of parameters is making Sturmian and eigenvalue problems equivalent. To
prove this

equivalence is nontrivial but possible. It was overlooked by D-O. In $[3]$
it is demonstrated

that even though Eq.s (3) and (4) are producing exactly the same spectrum,
only Eq.(4)

can be subjected to the conformal transformations while Eq.(3) cannot. That
such

transformations will lead to the correct reproduction of the multielectron
effects and are

complacent with the extended Bertrand theorem is also demonstrated in$[3]$.
The

complacency with the Bertrand theorem had become possible only thanks to
seminal work

by Volker Perlick $[23].$ In it the results of the classical Bertrand theorem%
$[20]$ valid in flat

Euclidean 3 dimensional space were extended to static spherically symmetric
spacetimes

of general relativity. By design, the motion in such curved spacetimes takes
place on

closed orbits. Thus, our task was to demonstrate that the
classical/semiclassical limit of

Eq.(4) with the appropriately deformed D-O potential leads to the motion in
generalized

Bertrand \ spacetimes found by Perlick. In$[3]$ such a demonstration was
performed. Thus,

for the first time the place of gravity effects in testable realistic
quantum mechanical

problem was found\footnote{%
More accurately, following J.A.Wheeler, we have to use the term
"geometrodynamics" instead of gravity. Recall, that Wheeler's
geometrodynamics is just elaboration on unified theory of gravity and
electromagnetism proposed by G.I.Rainich in 1925.}. In addition, in [$3$]
the connection between \ the

deformed D-O potential and the Hartree-Fock $V_{eff}(\mathbf{r}_{i})$
potential was found. These \ 

achievements enable us to make further progress\ described in this
work.\medskip \medskip

\textbf{3}. \textbf{Beyond the canonical Madelung rule \bigskip }

\textbf{3.1.\ The origin of the Madelung rule anomalies\medskip }

In the previous section we have defined the Madelung rule. The opposite of
this definition

can be taken as definition of Madelung-exceptions. Ref.[$15$] leaves us with
the impression

that the correct mathematical understanding of the empirical Madelung rule
can be made

only with uses of the results of the dynamical group theory while the D-O
results suggest

alternative approach which was significantly improved in [$3$]. If this is
so, is there in this

formalism a room for description of the Madelung-exceptional elements? From
section 1

\bigskip it follows that the Madelung exceptions are observed among \ some
transition metals,

lanthanides and actinides. The electronic structure of these elements was
studied thus far

with help of the relativistic Hartree-Fock methods [$24$]. The major new
problem emerges:

\textsl{will the results of solving Eq.(4) developed in detail in }$[3]$%
\textsl{\ \ survive the relativistic}

\textsl{extension}? Only such an extension may yield the results compatible
with that for the

Madelung-exceptional atoms. The most difficult issue in doing so is this. 
\textsl{Since the }

\textsl{already obtained nonrelativistic results are capable of deriving the
regular Madelung}

\textsl{rule quantum mechanically, the relativization of these results is \
going to make all }

\textsl{chemical elements anomalous since the} \textsl{standard }[$24$]%
\textsl{\ formalism works indiscriminantly}

\textsl{for all atoms}\textbf{.\medskip\ }The Seiberg-Witten (S-W) theory
helps to solve this puzzle. This can be

achieved in several steps.\medskip

\textbf{3.2. Preparing the nonrelativistic results for their relativistic
extension\medskip }

This \ extension can be achieved by using some not well known facts about
the quantization

of the hydrogen atom model Hamiltonian. These results will serve us as the
reference

point. In particular, in a specially chosen system of units the
dimensionless Hamiltonian

\^{H} \ for the hydrogen atom is given in the operator form as%
\begin{equation}
\text{\^{H}}=\mathbf{p}^{2}-\frac{2}{r}.  \tag{5}
\end{equation}

The Laplace-Runge-Lenz vector \textbf{A}$_{0}$ is given by%
\begin{equation}
\mathbf{A}_{0}=\frac{\mathbf{x}}{r}+\frac{1}{2}(\mathbf{L\times p-p\times L)}
\tag{6}
\end{equation}

with the angular momentum operator \textbf{L} defined as usual by \textbf{L}=%
\textbf{x}$\times \mathbf{p.}$ It is convenient

to normalize \textbf{A}$_{0}$ as follows%
\begin{equation}
\mathbf{A}=\left\{ 
\begin{array}{c}
\mathbf{A}_{0}(-H)^{\frac{1}{2}}\text{ for E\TEXTsymbol{<}0,} \\ 
\mathbf{A}_{0}\text{ for E=0,} \\ 
\mathbf{A}_{0}=(H)^{\frac{1}{2}},\text{ for E\TEXTsymbol{>}0.}%
\end{array}%
\right.  \tag{7}
\end{equation}

Here it is assumed that \^{H}$\Psi _{E}=$E$\Psi _{E}$ and E$=H.$ By
introducing two auxiliary angular

momenta \textbf{J}($\alpha ),\alpha =1,2,$ such that \ \textbf{J}($1)=\frac{1%
}{2}(\mathbf{L}+\mathbf{A})$ and \textbf{J}($2)=\frac{1}{2}(\mathbf{L}-%
\mathbf{A}),$and using

known commutation relations for \textbf{L}, etc., we arrive at%
\begin{eqnarray}
\mathbf{J}(\alpha )\times \mathbf{J}(\alpha ) &=&i\mathbf{J}(\alpha ),\alpha
=1,2,  \TCItag{8} \\
\lbrack \mathbf{J}(1),\mathbf{J}(2)] &=&0.  \notag
\end{eqnarray}

Taking into account that \textbf{L}$\cdot $\textbf{A}=0 we also obtain two
Casimir operators: \textbf{L}$\cdot $\textbf{A}=0=\textbf{A}$\cdot \mathbf{L}
$

and $\mathbf{L}^{2}+\mathbf{A}^{2}$. The Lie algebras $\mathbf{J}(\alpha
)\times \mathbf{J}(\alpha )=i\mathbf{J}(\alpha ),\alpha =1,2,$ are the
algebras of

rigid rotators for which the eigenvalues $j_{\alpha }(j_{\alpha }+1)$ are
known from the standard

texts on quantum mechanics. The peculiarity of the present case lies in the
fact that

$\mathbf{J}(1)^{2}=\mathbf{J}(2)^{2}$. This constraint is leading us to the
requirement: $j_{\alpha }=j_{\beta }=j.$ The

topological meaning of this requirement is explained in section 5 of [$3$].
\ In short,

the eigenvalue equation for the standard quantum mechanical rigid rotator is
that

for the Laplacian living on a 2-sphere $S^{2}$. \ Since in the present case
we are having

two rigid rotators, each of them should have its own sphere $S^{2}$.
However, the

constraint $j_{\alpha }=j_{\beta }=j$ causes these two spheres to be
identified with each other

pointwise. Topologically, such a pointwise identification leads to the
3-sphere $S^{3}.$

Group-theoretically the same result can be restated as $so(4)\simeq
so(3)\oplus so(3).$

With such background we are ready to relativize these results.\bigskip
\bigskip

\textbf{3.3. \ Sketch of derivation of the Madelung-regular rule\bigskip
\medskip }

To make sure that our relativization procedure is compatible with previously
obtained

results [$3$],we begin with restoration of these results in a new fashion\
using results of

previous subsection. For this purpose, the observation \ that the
3-dimensional rigid

rotator is having the eigenvalues and eigenfunctions of the Laplacian
"living on on $S^{2}$ 
\begin{equation}
\mathbf{L}^{2}Y_{lm}(\theta ,\phi )=l(l+1)Y_{lm}(\theta ,\phi ).  \tag{9}
\end{equation}

is helpful. Notice, however, that \textbf{L}$^{2}=$L$_{x}^{2}+$L$_{y}^{2}+$L$%
_{z}^{2}$ and L$_{x}=i$D$_{23},$L$_{y}=i$D$_{31},$L$_{z}=i$D$_{12}$ ,

where 
\begin{equation}
D_{\alpha \beta }=-x_{\alpha }\frac{\partial }{\partial x_{\beta }}+x_{\beta
}\frac{\partial }{\partial x_{\alpha }},\text{ \ \ \ }\alpha <\beta
=1,2,...d,  \tag{10}
\end{equation}

and $d$ is the dimensionality of space. Let now A$_{x}=iD_{14},$A$%
_{y}=iD_{24},$A$_{z}=iD_{34}.$

If $\mathbf{L}^{2}$ represents the Laplacian on $S^{2},$ the combination $%
\mathbf{L}^{2}+\mathbf{A}^{2}\equiv \mathcal{L}^{2}$ represents the

Laplacian \ on $S^{3}$ embedded in 4 d Euclidean space$[18].$ That is,
instead of more

familiar study of \ 3-dimensional rigid rotator "living" on the two-sphere $%
S^{2},$ the

eigenvalue problem for hydrogen atom is in fact involving the study of
spectrum of

the rigid rotator on $S^{3}$. This fact was realized initially by Fock [$25$%
]. The 3 Euler's

angles $\alpha ,\theta ,\phi $ on $S^{3}$ are replacing more familiar $%
\theta ,\phi $ angles used on the 2- sphere$.$

The eigenvalue, Eq.(9), is \ being replaced now by 
\begin{equation}
\mathcal{L}^{2}Y_{nlm}(\alpha ,\theta ,\phi )=I_{nl}Y_{nlm}(\alpha ,\theta
,\phi ).  \tag{11a}
\end{equation}

This result coincides with that obtained in the Appendix C, Eq.(C.1). Here
we have

the manifestly spherically symmetric wave functions with indices $n,l,m.$%
This result

is immediately applicable to the hydrogen atom [$18$]. It corresponds to the
choice

$\gamma =1$ in Eq.(\textbf{C5}). The choice $\gamma =1/2$ in the potential,
Eq.(B\textbf{.4}), results in the shift

in the indices in Eq.(11a) leading to%
\begin{equation}
\mathcal{L}^{2}Y_{n+l,lm}(\alpha ,\theta ,\phi )=I_{n+l,l}Y_{n+l,lm}(\alpha
,\theta ,\phi )  \tag{11b}
\end{equation}

in accord with qualitative arguments made in section 2. In spite of the
apparent

simplicity of transition from Eq.(11a) to (11b) and with account of results
of

Appendices B and C, lengthy calculations $[3]$ are still required. For the

hydrogen atom the spectrum associated with Eq.(11a) is obtained in the
Appendix C,

Eq.(C.3). While for the multielectron atom obeying the regular Madelung
rule, the

spectrum associated with Eq.(11b) is given below, in Eq.(27). Now we are in
the

position to develop the theory explaining the Madelung-exceptional
atoms.\medskip

\textbf{3.4. \ Uncovering the source of the Madelung rule exceptions via}

\ \ \ \ \ \ \ \ \textbf{relativization of results of previous subsection
\medskip \medskip }

This task can be completed in several steps. First, we notice that in the
standard 3

dimensional calculations the hydrogen spectrum is determined by the
eigenvalues

of the \textsl{radial equation}%
\begin{equation}
\lbrack -\frac{1}{2}(\frac{d^{2}}{dr^{2}}+\frac{2}{r}\frac{d}{dr}-\frac{%
l(l+1)}{r^{2}})+V(r)]R_{El}(r)=ER_{El}(r).  \tag{12}
\end{equation}

Here, the total wave function

$\Psi _{E}=F_{El}(r)\mathcal{Y}_{lm}(\theta ,\phi ),\mathcal{Y}_{lm}(\theta
,\phi )=r^{l}Y_{lm}(\theta ,\phi ),R_{El}(r)=r^{l}F_{El}(r)$

and $V(r)=-\frac{Ze^{2}}{r},m=1,\hbar =1.$ The combination $F_{El}(r)%
\mathcal{Y}_{lm}(\theta ,\phi )$ can be rewritten

in terms \ of $Y_{nlm}(\alpha ,\theta ,\phi )$ as demonstrated in$[3]$ in
accord with [$18$]. Therefore,

it is sufficient to look at 3 dimensional results. They can always be mapped

into $S^{3}$ via inverse stereographic projection. Next, this observation
allows us, following

Martin and Glauber$[26]$ and Biedenharn [$27$] to use the Pauli matrices $%
\sigma _{i}$ in order to

rewrite $\mathbf{L}^{2}=\left( \mathbf{\sigma }\cdot \mathbf{L}\right) 
\mathbf{(\sigma }\cdot \mathbf{L+}1).$ This identity permits us then to
write the total

momentum $\mathbf{J}$ as $\mathbf{J}=\mathbf{L}+\frac{1}{2}\mathbf{\sigma }$%
. After that, it is convenient to introduce the

operator $\mathcal{K=}\mathbf{\sigma }\cdot \mathbf{L+}1$ used already by
Dirac$[28]$ in his treatment of hydrogen atom

with help of the Dirac equation. Using \ this operator it is possible to
obtain the

identity: $\mathcal{K}^{2}=\mathbf{J}^{2}+\frac{1}{4},\hbar =1.$ The
eigenvalues of $\mathcal{K}$,\ denoted as $\kappa $, are known

to be $\kappa =\pm 1,\pm 2,..(0$ is excluded$).$Use of these results implies:%
\begin{eqnarray}
l &=&l(\kappa )=\left\{ 
\begin{array}{c}
\kappa ,\text{ if }\kappa \text{ is positive} \\ 
\left\vert \kappa \right\vert -1,\text{ if }\kappa \text{ is negative}%
\end{array}%
\right\vert  \notag \\
j &=&j(\kappa )=\left\vert \kappa \right\vert -\frac{1}{2}.  \TCItag{13}
\end{eqnarray}

The above results were presented with purpose not at all discussed in the

standard texts on quantum mechanics. Specifically, at the classical level,
the

Kepler trajectories can be determined with help of the vector \textbf{A }%
only[$29$]. This

fact suggests that the quantum analog of \textbf{A} should produce the
eigenvalue

spectrum identical to that obtained using Eq.(12). This is indeed the case.

To demonstrate this, we introduce the operator $\mathcal{N}$ such that $%
\left( \mathcal{N}\right) ^{2}=$

$\left( \mathbf{\sigma }\cdot \mathbf{A}\right) ^{2}$ +$\left( \mathcal{K}%
\right) ^{2}.$ Since it can be shown that $\mathbf{\sigma }\cdot \mathbf{A}$
and $\mathcal{K}$ anticommute,

it becomes also possible to write%
\begin{equation}
\mathcal{N}=\mathbf{\sigma }\cdot \mathbf{A}+\mathcal{K}.  \tag{14}
\end{equation}

Denote the eigenvalues of $\mathcal{N}$ as $\pm N.$ Then, it is possible to
demonstrate that 
\begin{equation}
\mathbf{\sigma }\cdot \mathbf{A\mid }N,\varkappa ,m>=(N^{2}-\varkappa ^{2})^{%
\frac{1}{2}}\mid N,-\varkappa ,m>.  \tag{15}
\end{equation}

It is possible as well to demonstrate that $N\rightleftarrows E$ with $E$
defined in Eq.(12).With

help of this result it is possible next to write the exact equivalent of the
radial

Eq.(12). It is given by 
\begin{equation}
\lbrack \frac{1}{r^{2}}\frac{d}{dr}r^{2}\frac{d}{dr}-\frac{\mathcal{K}(%
\mathcal{K}+1)}{r^{2}}+\frac{2Ze^{2}}{r}-k^{2}]F_{N,l(\kappa )}(r)=0. 
\tag{16}
\end{equation}

Here $k^{2}=2\left\vert E\right\vert ,m=1,\hbar =1.$ Biedenharn [$27$]
explains how the wave function

$\mid N,-\varkappa ,m>$ \ is related to $F_{N,l(\kappa )}(r).$ Also, $%
\mathcal{K}(\mathcal{K}+1)=l(\kappa )(l(\kappa )+1).$ \bigskip

\textsf{Not only just presented results demonstrate that the quantum version}

\textsf{of the \ Laplace-Runge-Lenz operator leads to the eigenvalue problem}

\textsf{identical to the standard eigenvalue problem, Eq.(13), for hydrogen }

\textsf{atom presented in every textbook on quantum mechanics but, in
addition,}

\textsf{these results permit us to perform their relativization the most
naturally }

\textsf{thus allowing seamless match of relativistic results with those
known }

\textsf{from the nonrelativistic quantum mechanics.}

The control parameter of this relativistic generalization is the fine
structure constant

$\alpha =\frac{e^{2}}{c\hbar }.$ In the limit $\alpha =0$ \ the result,
Eq.(16), is recovered as required. Since

structurally it is identical with Eq.(12), the nonrelativistic spectrum is
preserved.

For $\alpha >0$ Eq.(17) is replaced by a very similarly looking equation 
\begin{equation}
\lbrack \frac{1}{r^{2}}\frac{d}{dr}r^{2}\frac{d}{dr}-\frac{\Gamma (\Gamma +1)%
}{r^{2}}+\frac{2\alpha ZE}{c\hbar r}-k^{2}]\Phi _{N,l(\gamma \kappa )}(r)=0.
\tag{17}
\end{equation}

Here, to avoid confusion, when comparing with the original sources, we
restore $\hbar ,c$

and $m.$ In particular, $k^{2}=[\left( m^{2}c^{4}-E^{2}\right) /c^{2}\hbar
^{2}],\Gamma $ is the Lippmann-Johnson operator%
\begin{equation}
\Gamma =\mathcal{K}+i\alpha Z\rho _{1}\mathbf{\sigma }\cdot \mathbf{\check{r}%
},  \tag{18}
\end{equation}

$\mathbf{\check{r}}=\frac{\mathbf{x}}{r},\rho _{1}\div \rho _{3},\sigma
_{1}\div \sigma _{3}$ are the $4\times 4$ matrices defined in Dirac's book $%
[28]$. Instead

of the eigenvalue $\kappa $ for $\mathcal{K}$ now one has to use $\gamma
\kappa $ so that, upon diagonalization,

$\Gamma (\Gamma +1)=l(\gamma \kappa )(l(\gamma \kappa )+1)$ and%
\begin{equation}
l(\gamma \kappa )=\left\{ 
\begin{array}{c}
\gamma \kappa =\left\vert \kappa ^{2}-\left( \alpha Z\right) ^{2}\right\vert
^{\frac{1}{2}}\text{ for }\gamma \kappa >0 \\ 
\left\vert \gamma \kappa \right\vert -1=\left\vert \kappa ^{2}-\left( \alpha
Z\right) ^{2}\right\vert ^{\frac{1}{2}}-1\text{ for }\gamma \kappa <0%
\end{array}%
\right. .  \tag{19}
\end{equation}

Mathematically, both Eq.s(16) and (17) are looking almost the same and, in
fact,

their solution can be reconstructed from solution of the radial eigenvalue

Eq.(12) discussed in any book on quantum mechanics.\medskip\ Details are
given in

the Appendix A\medskip .

\textsf{The difference between these equations lies only in redefining the }

\textsf{parameter }$l:$\textsf{\ In the nonrelativistic case the combination 
}$l(\kappa )(l(\kappa )+1)$

\textsf{is the same as }$l$\textsf{(}$l$\textsf{+1) as required, while in
the relativistic case we}

\textsf{should replace }$l$\textsf{\ by }$l(\gamma \kappa ).$ \textsf{By
replacing }$l$\textsf{\ in Eq.(12b) by }$l(\gamma \kappa )$\textsf{\ it is }

\textsf{immediately clear that the Madelung rule in its canonical formulation%
}

\textsf{is no longer valid.\bigskip \bigskip }

\textbf{4}. \ \textbf{New physics behind the Madelung rule anomalies\
\bigskip }

\textbf{4.1.} \ \textbf{The}\ \textbf{Madelung rule and its anomalies
explained with help of}

\ \ \ \ \ \ \ \ \textbf{Schr\"{o}dinger's work on Dirac electron in a
gravitational field\medskip \bigskip\ \medskip }

In 1932 the paper by Schr\"{o}dinger $[30]$ on Dirac electron in a
gravitational field

was published. Historically, Dirac $[28]$came up with his equation in 1928
being

driven by the observation that the Schr\"{o}dinger equation is not Lorentz
invariant.

By correcting this deficiency Dirac uncovered the spin of electron in 1928.

In 1927 the spin was artificially inserted into Schr\"{o}dinger's equation
by Pauli.

Schr\"{o}dinger immediately got interested in Dirac's equation and wanted to
study

how \ Dirac's formalism might be affected by gravity. The rationale for
doing so

is \ given in Schr\"{o}dinger's paper. Modern viewpoint will be presented
below. In

this subsection we discuss Schr\"{o}dinger's results in the light of their
relevance to

the description of Madelung rule and its exceptions in view of the noticed

relevance of \ Volker Perlick's work$[23]$ on the generalized Bertrand
theorem to

spectral problems of the atomic physics. In $[3]$, the generalized Bertrand
theorem

was used for the derivation of the regular Madelung rule. To explain the
exceptions

we need to relativize the already presented calculations. This process was
initiated

in section 3 and Appendix A.

\ \ \ We begin with the Dirac equation 
\begin{equation}
i\gamma ^{a}\partial _{a}\psi -m\psi =0  \tag{20a}
\end{equation}

in which the Dirac gamma matrices $\gamma ^{a}$ obey the Clifford algebra
anticommutation

rule : $\gamma ^{a}\gamma ^{b}+\gamma ^{b}\gamma ^{a}=2\eta ^{ab},$ $%
a,b=1\div 4,\eta ^{ab}$ is the matrix enforcing the Minkowski

spacetime signature $\{1,-1,-1,-1\}$. As is well known, the equivalence
principle

of general relativity locally allows to eliminate the effects of gravity
(e.g. recall

the falling elevator gedanken experiment). Mathematically, this can be
achieved

by introduction of a vierbein $e_{\mu }^{a}(x)$ so that $e_{\mu
}^{a}(x)e_{\nu }^{b}(x)\eta _{ab}=g_{\mu \nu }(x)$ and

$e_{a}^{\mu }(x)e_{b}^{\nu }(x)g_{\mu \nu }=\eta _{ab}(x).$ Thus, the
vierbeins carry in themselves the effects of

gravity since the metric tensor $g_{\mu \nu }(x)$ carries the information
about gravity.

Introducing these effects into Eq.(20a) can be done as follows. First, the

anticommutator $\gamma ^{a}\gamma ^{b}+\gamma ^{b}\gamma ^{a}=2\eta ^{ab}$
is replaced by $\gamma ^{\mu }\gamma ^{\nu }+\gamma ^{\nu }\gamma ^{\mu
}=2g^{\mu \nu }$ with

help of the relationship $\gamma ^{\mu }=e_{a}^{\mu }\gamma ^{a}.$ Here the
Greek indices $\mu $ and $\nu $ refer to the

4-dimensional spacetime while the Latin indices $a,b$ \ are referring to the

Lorentzian (more generally, to the Poincar$e^{\prime }$) frames. The
Lorentzian frames are

used for description of rotations in 4 dimensional spacetime of special
relativity

while the Poincar$e^{\prime }$frames account for translations in addition.
The partial

derivative $\partial _{\mu }$ is replaced \ \ now by the covariant derivative%
\begin{equation}
\nabla _{\mu }\psi =\partial _{\mu }\psi +\Gamma _{\mu }\psi ,  \tag{21}
\end{equation}

where 
\begin{equation}
\Gamma _{\mu }(x)=-\frac{i}{4}\omega _{ab\mu }(x)\sigma ^{ab};\sigma ^{ab}=%
\frac{i}{2}[\gamma ^{a},\gamma ^{b}]  \tag{22}
\end{equation}

and 
\begin{equation}
\omega _{b\mu }^{a}=e_{\nu }^{a}\partial _{\mu }e_{b}^{\nu }+e_{\nu
}^{a}e_{b}^{\rho }\Gamma _{\rho \mu }^{\nu }.  \tag{23}
\end{equation}

In the simplest case $\Gamma _{\rho \mu }^{\nu }$\ is the standard
Levi-Civita connection determined by the

metric tensor $g_{\mu \nu }.$ Presence of the term $e_{\nu }^{a}\partial
_{\mu }e_{b}^{\nu }$ in Eq.(23) is responsible for the

torsion effects. These are absent in the canonical general relativity.
Extension of

general relativity accounting for the torsion effects is known as the
Einstein-Cartan

(ECG) gravity [$31$]. Use of Eq.(21) converts the flat space Dirac Eq.(20a)
into that

in the curved space%
\begin{equation}
i\gamma ^{\mu }\nabla _{\mu }\psi -m\psi =0.  \tag{20b}
\end{equation}

Instead of Eq.(20b) we can consider the following equation:%
\begin{eqnarray}
0 &=&(-i\gamma ^{\mu }\nabla _{\mu }\psi -m\psi )(i\gamma ^{\nu }\nabla
_{\nu }\psi -m\psi )  \notag \\
&=&\gamma ^{\mu }\gamma ^{\nu }(\nabla _{\mu }\nabla _{\nu }+\nabla _{\nu
}\nabla _{\mu }+\nabla _{\mu }\nabla _{\nu }-\nabla _{\nu }\nabla _{\mu
}+m^{2})\psi  \notag \\
&=&(g^{\mu \nu }\nabla _{\mu }\nabla _{\nu }+m^{2}+\frac{1}{8}R_{\alpha
\beta \delta \eta }\gamma ^{\mu }\gamma ^{\nu }\gamma ^{\delta }\gamma
^{\eta })\psi ,  \TCItag{20c}
\end{eqnarray}

where the following identity was used $[32]$ :%
\begin{equation}
\left( \nabla _{\alpha }\nabla _{\beta }-\nabla _{\beta }\nabla _{\alpha
}\right) \psi =\frac{1}{8}R_{\alpha \beta \delta \eta }\gamma ^{\delta
}\gamma ^{\eta }\psi  \tag{24}
\end{equation}

along with the Clifford algebra anticommutator identity $\gamma ^{\mu
}\gamma ^{\nu }+\gamma ^{\nu }\gamma ^{\mu }=2g^{\mu \nu }.$

Here $R_{\alpha \beta \delta \eta }$ is the Riemannian curvature tensor.

\ \ The above equation can be rearranged further$[33]$ yielding the
equivalent

final result:%
\begin{equation}
\left( g^{\mu \nu }\nabla _{\mu }\nabla _{\nu }+m^{2}+\frac{R}{4}\right)
\psi =0.  \tag{20d}
\end{equation}

Here $R$ is the scalar curvature. As it is demonstrated in$[3],[34]$ the
mass term $m^{2}$

is not essential and can be eliminated by the appropriate substitutions. In
the

case of fish-eye potential this is discussed also in Appendix B. In$[3]$,

sections 3 and 5, it is demonstrated \ that Eq.(20d) (with $m=0$) is \textbf{%
exactly}

\textbf{equivalent} to Eq.(4). In the mathematical literature such an
equation is known

as one of the Weitzenbock-Lichnerowicz equations [$33,35$], another example
is given

in Eq.(25) below. Additional information is presented in the Appendix E.2.

These type of equations are discussed further below in the context of the
Seiberg

-Witten theory. The scalar curvature $R$ in Eq.(20d) can be identified

with the \ potential, Eq. (B.4\textbf{)}, (with $\gamma =1/2),$e.g. read the
Appendix B$.$ That such

chosen scalar curvature coincides with the curvature of the Bertrand space
was

demonstrated in $[3]$ and, independently, in $[36]$. The obtained result,
Eq.(20d),

is incomplete though. To make it complete, following Schr\"{o}dinger [$30$]
we have

to modify the definition of the covariant derivative in Eq.(21). That is we
have

to replace $\nabla _{\mu }=\partial _{\mu }+\Gamma _{\mu }$ by $\nabla
_{A\mu }=\partial _{\mu }+\Gamma _{\mu }-ieA_{\mu }$,where $A_{\mu }$ is
some kind of a

vector (e.g. electromagnetic)

potential. With such a replacement, \ Eq.(20d) is replaced now by 
\begin{equation}
\left( g^{\mu \nu }\nabla _{\mu }\nabla _{\nu }+m^{2}+\frac{R}{4}+\frac{ie}{2%
}\sigma ^{ab}F(A)_{ab}\right) \psi =0;F(A)_{ab}=\partial _{a}A_{b}-\partial
_{b}A_{a}.  \tag{25}
\end{equation}

This is the final result obtained by Schr\"{o}dinger. In the mathematical
literature the same

equation also known as Lichnerowicz-Weitzenbock equation (more on this is
presented

in the Appendix E.2). Most of calculations in S-W theory involve uses of
this equation

[$37$]. The signs of $i$ and $e$ factors in Eq.(25) can be correctly
restored. For this we have

to put $\Gamma _{\mu }=0$ in the covariant derivative $\nabla _{\mu }$ and
then, to consult the book on quantum \ 

electrodynamics, e.g., see [$38$]$,$ page 66, Eq.(2.73).

\ \ \ Eq.s(16) and (17) now can be related to Eq.s(20d) and (25).
Specifically, by putting

the fine structure constant $\alpha $ in the Lippmann-Johnson operator to
zero we are arriving

at Eq.(20d). For the nonzero $\alpha $ we have to use Eq.(25) instead.
Important details are

presented below. This kind of logic, common in physics literature [$38]$,
does not take

into account finer details, e.g. the topological considerations, etc. In the
next subsection

we initiate the discussion of this topic.\bigskip \medskip

\textbf{4.2. The topological transition between the Madelung-regular and
Madelung-}

\ \ \ \ \ \ \textbf{exceptional atoms. Motivation\medskip }

\ \ \ \ \ Just \ presented results permit us now to explain \ the origins of
the Madelung rule

exceptions mainly using arguments \ familiar from physics literature. This
explanation

and the results of Appendices E and F will provide an answer to the question:

\textsf{Why for the most elements the relativistic effects are negligible
and why, without}

\textsf{exceptions, they are significant for description of atoms exhibiting
the Madelung rule }

\textsf{exceptions}? To proceed, it is helpful to \ make several additional
comments.

1. The term $\frac{ie}{2}\sigma ^{ab}F_{ab}$ in Eq.(25) is responsible for
the relativistic effects. Without this

term Eq.(25) is converted into Eq.(4) in which $V_{eff}(\mathbf{r})$ is
represented by $\dfrac{R}{4}$. By identifying

Eq.s (4) and (21d) we must identify $\dfrac{R}{4}$ with $V(r)$, Eq.(B.4), in
which $\gamma =1/2.$

2.The relativistic Eq.s (17) and (25) are equivalent even though the
mathematicians

prefer to work with Eq.(25) for deep reasons to be explained below.

3.The relativistic Eq.(17) and nonrelativistic Eq.(16) look almost the same.
Thanks to

the work$[39],$ both equations can be made to look \textbf{exactly} the same
(up to the difference

in the meaning of constants in these equations). This is demonstrated in the

Appendix A.

\textsl{Because of this circumstance, all results obtained in }$[3]$ \textsl{%
for the nonrelativistic \ case }

\textsl{can now be transferred to the relativistic case unchanged.}

4. It is of interest to derive the spectrum of the Dirac-Coulomb problem by
using methods

developed in the Appendix A via replacing the Coulomb potential with the
fish-eye

potential first and then, by applying the Wong and Yeh results$[39]$. This
is done in

Appendices B and C.

\ \ \ \ In the Appendix B we demonstrate the equivalence of the Coulomb and
fish-eye

classical and quantum problems while in the Appendix C we demonstrate how the

hydrogen atom spectrum (both non and relativistic) can be obtained using the
fish-eye

potential. The treatment of multi-electron atoms with help of the deformed
fish-eye

potential, Eq.(B.4), then follows the same steps as outlined in the Appendix
C.

Additional details are given in Ref.[$3$].

\ \ \ \ \ With such a background, following[$38$]$,$page 75, we can now
write down the fine

structure

$\alpha $ expansion for the spectrum of hydrogen atom ($\hbar =1)$ 
\begin{equation}
E_{nj}=mc^{2}-\frac{mc^{2}\alpha ^{2}Z^{2}}{2\tilde{n}^{2}}-\frac{\alpha
^{4}Z^{4}mc^{2}}{2\tilde{n}^{4}}[\frac{n}{j+\frac{1}{2}}-\frac{3}{4}%
]+O(\alpha ^{6}).  \tag{26}
\end{equation}

Here $\tilde{n}=n_{r}+l+1$ in accord with results of Appendix C, $j=\frac{1}{%
2},\frac{3}{2},...,\tilde{n}-\frac{1}{2}.$

\textsf{The }$\alpha ^{4}$\textsf{\ relativistic correction is in fact
coming directly from the spin orbital }

\textsf{interaction [38],} pages 73-75. Since this fact is well known but of
fundamental

importance for this paper, it will be discussed in detail below within the

context of S-W theory. In the meantime, by using the deformed

fish-eye potential, Eq.(B.4) ( $\gamma =1/2),$ and, by repeating
calculations described in

the Appendices B and C using such deformed potential, the limit $\alpha =0$
is obtained.

To achieve it requires replacing $\tilde{n}$ by $\tilde{n}+l$ in Eq.(C.3),
thus resulting in the regular

Madelung rule ($\hbar =1)$: 
\begin{equation}
E_{nl}=mc^{2}-\frac{me^{2}Z^{2}}{2\left( \tilde{n}+l\right) ^{2}}.  \tag{27}
\end{equation}

For $\alpha \neq 0$ a simple minded use of the expansion, Eq.$(26$), in Eq.($%
27$) leads to

the entirely wrong results. The relativistic effects cannot be neglected
already

for the hydrogen atom as is well known. They are responsible for the fine

structure spectrum. Simple minded application of the same logic to the
multielectron

atoms then leads us to the conclusion that, \textbf{based on methods
utilized thus }

\textbf{far, treatment of the Madelung-exceptional atoms cannot acquire
special}

\textbf{status. Either all atoms should obey the standard Madelung rule}

\textbf{or they all should be exceptional.}

The resolution of this paradoxical situation is obtained by providing
ramifications

to the item 2 (see above). That is now we need to explain why mathematicians

prefer to work with Eq.(25) instead of Eq.(17). Very deep results in
mathematics

based on the theory of fiber bundles and spinor bundles, e.g. read $[35],$%
pages 152-154,

or$[33]$, pages 40-41, imply that Eq.(21d) makes sense only for \textit{spin}
manifolds.\textbf{\ That}

\textbf{is the quantum mechanical description of atoms exhibiting the
canonical }

\textbf{Madelung-regular behavior should be described exclusively \ using
Eq.(20d)}.

The account of the relativistic fine structure effects is made with help of
Eq.(25). This

equation is living on the \textit{spin}$^{c}$ -type manifolds[$33,35,40$]. 
\textbf{The description of the }

\textbf{transition- from\ atoms obeying the regular to atoms obeying the
exceptional}

\textbf{Madelung rule-cannot\ be achieved with help of known \
perturbational }

\textbf{methods since it is topological in nature. It is topological since
the }\textit{spin}\textbf{\ and }

\textit{spin}$^{c}$\textbf{\ \ are \ topologically different manifolds. They
cannot be smoothly }

\textbf{transformed into each other.\medskip }

This circumstance will be explained in detail below. Before doing so,

we would like to explain the difference between the \textit{spin}\textbf{\ }%
and\textbf{\ }\textit{spin}$^{c}$\ manifolds using

terminology\ familiar to chemists and physicists. More mathematically
rigorous results

are presented in appendices culminating in Appendix E.4.\ The differences
between

these manifolds is in the differences in underlying spin symmetry. The
topology and

symmetry are intertwined as is well known. More specifically, the results
depend

upon wether the underlying manifold M is being orientable or nonorientabe.

Topologically, this is described in terms of the 1st Stiefel-Whitney class w$%
_{1}$(M). For

the orientable manifold w$_{1}$(M)=0. If, in addition, the manifold can
carry the \textit{spin}

structure (Appendix D), the second Stiefel-Whitney class w$_{2}$(M) also
should be zero.

This is beautifully explained in Nakahara's book$[41]$, pages 404, 405.

The manifold M which carry $spin^{c}$ is described in terms of the
requirement$[42],$page 123,

c$_{1}$(M)=w$_{2}$(M) $\func{mod}2.$ Here c$_{1}$(M) is the 1st Chern class.
For the spin manifolds w$_{2}$(M)=0

and this relation breaks down. This explains why the $\mathit{spin}$ and $%
spin^{c}$are topologically

different manifolds. However, such an explanation it is not describing the
underlying

physics well. The underlying physics is explained in Appendices E and F.

Locally, in the language familiar to physicists, it is sufficient to look at
the

the difference between, say, the set of compass arrows (dipoles),
emphasizing the

orientability of space in which they live, and the set of nematic molecules
(dipoles

without charges at the ends), \ emphasizing the nonorientability of the
projective spaces

which they suppose to illustrate. It is instructive to check if such
simplified physical

description \ can \ be used for \ visualization of $\mathit{spin}$ and $%
spin^{c}$\ structures. \ In Appendix F

we are checking if such very simplified description of the $\mathit{spin}$
and $spin^{c}$\ structures,

when applied to the Madelung-exceptional atoms, makes sense. Such simplified

treatment should be considered only as qualitative/nonrigorous. The rigorous

treatment in presented in Appendix E.

\ \ \ \ With these remarks behind us, we still have to demonstrate the
equivalence

of \ Eq.s(17) and (25) in order to demonstrate that the relativistic
corrections

are coming (in part) from the spin-orbital interactions. Ramifications of
such

demonstration will link the Hund rule, the LS, JJ and LSJ couplings schemes

to the\textit{\ \ spin} $\rightleftarrows $ \textit{spin}$^{c}$ topological
transition.

Following [$38$], page 74, \ without loss of generality we have (for the

centrally symmetric Coulombic field):%
\begin{equation}
\frac{ie}{2}\sigma ^{ab}F_{ab}=\pm ie\mathbf{\sigma }\cdot \mathbf{E}=\pm
iZ\alpha \frac{\mathbf{\sigma }\cdot \mathbf{\check{r}}}{r^{2}},  \tag{28}
\end{equation}

where $\mathbf{\check{r}}$ is the unit vector. A quick look at
Eq.s.(17),(18) allows us to realize that just

obtained result enters into the Lippman -Johnson operator, Eq.(18).Thus, \
at least for

the case of single electron, \ Eq.s (17) and (25) do coincide. And if this
is so, by applying

the Foldy-Wouthuysen \ transformation to Eq.(17) we obtain, \textbf{in the
first order in} $\alpha ,$

the spin-orbit coupling interaction term[38],pages 69-75. This means that
Eq.(17)

contains information on spin-orbit coupling \textsl{to all orders in} $%
\alpha $ and, therefore, spares

us from adding the spin-orbital correction to the nonrelativistic
Hamiltonian. The

situation with this term becomes more complicated for multielectron atoms.
Detailed

calculations presented in the series of three papers by Blume and Watson[$%
46,47$].

These papers and references therein indicate that, very fortunately, the
complicated

expressions can be squeezed back to the spin-orbital interaction Hamiltonian
known

for the hydrogen atom with the appropriately redefined coupling constant to
be

determined experimentally. This fact is not affecting the analytical
structure of Eq.(17)

and, therefore, the \textbf{exact} mapping- from the relativistic to
nonrelativitic case-described

in the Appendix A remains intact. Since the spin-orbit interaction
Hamiltonian is the

first order in $\alpha $ result, it surely cannot compete with the
topological arguments [$35$],

pages 152-154, or [$33$], pages 40-41, valid to all orders in $\alpha .$
Just presented facts, and

those in Appendix F, explain the nature of the Madelung rule anomalies at
the physical

level of rigor.\bigskip \medskip

\textbf{5. \ Using Seiberg-Witten theory for explanation of the normal to }

\ \ \ \ \textbf{superconducting topological transition at the atomic
level\medskip }

\bigskip

Already in subsection \ 1.1. we noticed that the diamagnetic properties of
hydrogen atom

are making it superconductor at the atomic level. None of other hydrogen
-like atoms ,

e.g. Li,Na,K, Rb,Cs are diamagnetic. In fact, they are all paramagnetic. At
the

same time, not at all surprising, all noble gases are diamagnetic. The
hypothesis by

Ashcroft and Ginzburg about hydrogen superconducticity brings along the
following

questions. 1. If the atomic hydrogen is a superconductor (since it is
subject to the

Meissner effect), can we call the noble gases as superconductors? 2.If
condensed

and solidified, will all these elements become superconducting? 3. Under the

appropriate pressure-temperature \ conditions, will all these elements
become at

least conductors? 4. Are there properties, other than zero resistivity,
indicating

that a given solid is superconductor? Some answers to questions 1-3 can be
found

in [$5$]. As for question 4, we would like to mention the following. \
Historically, the

Madelung-exceptional palladium is the first element allowing us to find an
answer

to the question 4. In subsection 1.2. it is stated that Pd is not
superconductor but PdH$_{x}$

is. Here, as in section 1.2., we are talking first about PdH$_{x}$ under
ambient pressure.

Negligibly small amounts of the absorbed hydrogen make palladium
superconducting

and its T$_{c}$ \ is going up directly proportional to the amount of
absorbed H. Since the

absorption is reversible (that is, it costs zero energy) palladium is
fantastic catalyst and

hydrogen storage provider[$11$]. The property of reversible absorption
allows to apply

to solid palladium the method of quasiaverages developed by Bogoliubov [$13$%
],[$14$].

Although eventually this method was applied to many order-disorder phase
transitions,

initially it was applied to the BCS superconductors. Method of quasiaverages
explains

why the superconducting condensate is not number-conserving. That is, the
number of

Cooper pairs in superconducting condensate is not conserved. \ Below we
argue that:

a) all Madelung-exceptional atoms are superconductors in the sense of
conventional

mathematical description of superconductivity done either within a framework
of the

Ginzburg-Landau or, more sophisticated, Seiberg-Witten theory; b) this
property at

the level of individual atoms survives solidification due to experimentally
observed

reversible absorption- the quality shared by all Madelung-exceptional
elements [$11$].

\ \ \ \ To demonstrate a), it is instructive to reconsider \ the equivalence
between Eq.s(17)

and (25) first. Thus far we used plausible arguments following book by
Itzykson and

Zuber [$38$], page 74. These plausible arguments can be made rigorous using
results

of S-W formalism. For the sake of space, we expect our readers to have some

familiarity with this formalism at least at the level of books by Jost$[35]$
and Naber [$40$].

To expedite matters we also recommend reading, \ at least the first couple
pages, of the

review by Donaldson [$38$].

\ \ \ \ To make our first step, that is to restore Eq.(28) using the S-W
equations, we shall

follow the paper by Naber[$40,49$]. Using this paper it is sufficient to
consider these

equations in the flat Minkowski spacetime \textbf{R}$^{1,3}.$In such a case
the covariant derivative

$\nabla _{A\mu }=\partial _{\mu }+\Gamma _{\mu }-ieA_{\mu }$ should have $%
\Gamma _{\mu }=0$. For complacency with [$40,49$] we rewrite

$\nabla _{\mu }$ as $\nabla _{\mu }=\partial _{\mu }+A_{\mu }.$ Then, the
first of \ S-W equations can be written as

\begin{equation}
\gamma ^{\mu }\nabla _{A\mu }\psi =0.  \tag{29}
\end{equation}

To write the second of S-W equations, it is essential to keep in mind the
origin of

these equations. \ At first look, it appears that it is sufficient to
consult the book of Jost

$[35]$ or, the paper by Witten$[50].$ From both sources it follows that the
S-W

equations emerge as generalization of the G-L equations of
superconductivity. Thus,

the solutions of S-W equations must contain vortices/monopoles-typical
solutions

of G-L equations. At the very advanced level, this fact was reconfirmed by
Taubes

[$51$]. In the present case our Eq.(25), although included into the S-W
formalism,

requires some additional explanations. These are presented in the Appendix
E, thus

making our atomic physics problem an intrinsic part of the S-W formalism.

\ \ \ At the same time, the treatment of G-L equations begins typically with
writing of

the G-L functional whose variation is producing the set of G-L equations.
The self-

duality considerations allow then to simplify calculations considerably and
to

reduce the order of these equations-from 2 to1. Very detailed exposition of
this topic

is given in our book$[52],$ chapters 5.6. Such a reduction was performed
first in the

context of the Yang -Mills fields by Bogomolny whose methodology was extended

to S-W theory, where the 1st order self-dual equations are also used.
Naber's paper

[$49$] also uses this reduction. Thus obtained S-W solution, even though
reproduces the

result, Eq.(28), is not $L^{2}$ normalizable. Similar cases of $L^{2}$
normalizable solutions

are discussed in[$53$]. For the S-W monopoles on K\"{a}hler and symplectic
manifolds

more advanced treatment is presented in $[54].$ These lecture notes use
essentially

the results of Taubes$[51]$.

\ \ \ \ \ With these remarks we return back to Eq.(25) without mass term. It
can be

eliminated as it was explained in[$3,34$]. Let $\mathcal{D}_{A}$= $\gamma
^{\mu }\nabla _{A\mu }$ then, using the massless

Eq.(25), we obtain: 
\begin{equation}
\dint\limits_{M}\left( \mathcal{D}_{A}^{+}\bar{\psi},\mathcal{D}_{A}\psi
\right) dvol=\dint\limits_{M}\{(\nabla _{A}^{+}\bar{\psi},\nabla _{A}\psi
)^{2}+\frac{R}{4}(\bar{\psi},\psi )+\frac{1}{2}(F^{+}(A)\bar{\psi},\psi
)\}dvol=0,  \tag{30}
\end{equation}

where \ we used the Hermitian scalar product (,), and the self-dual portion
of the $F(A),$

that is $F(A)^{+}=\frac{1}{2}(F(A)+\ast F(A)).$ The notion of \textsl{spinor
bundle} (Appendix D) allowed

us to write $F(A)\psi =F(A)^{+}\psi ,$ e.g. read page 76 of Ref.[$55]$.

The result, Eq.(30), should be compared against the standard G-L functional:%
\begin{equation}
S_{G-L}(A,\psi )=\dint\limits_{M}\{(\nabla _{A}\psi )^{2}+\left\vert
F(A)^{+}\right\vert ^{2}+\frac{R}{4}\left\vert \psi \right\vert ^{2}+\frac{1%
}{8}\left\vert \psi \right\vert ^{4}\}dvol.  \tag{31}
\end{equation}

To make Eq.s(30) and (31) to coincide formally, following Jost$[35],$ we
need:

a) to write the 2-form $F(A)^{+}$as $F(A)^{+}=F(A)_{ij}^{+}\gamma ^{i}\wedge
\gamma ^{j},$b) to assume that

$F(A)_{ij}^{+}=\frac{1}{4}(\gamma _{i}\cdot \gamma _{j}\psi ,\psi ).$Here
the dot$\cdot $ represents the Clifford multiplication

(Appendix D). After that, we formally obtain:%
\begin{equation}
S_{S-W}(A,\psi )=\dint\limits_{M}\{\left( \mathcal{D}_{A}\psi \right)
^{2}+\left\vert F(A)^{+}-\frac{1}{4}(\gamma _{i}\cdot \gamma _{j}\psi ,\psi
)\gamma ^{i}\wedge \gamma ^{j}\right\vert ^{2}\}dvol=S_{G-L}(A,\psi ) 
\tag{32}
\end{equation}

leading to the 1st, Eq.(29), and to the 2nd%
\begin{equation}
F(A)^{+}=\frac{1}{4}(\gamma _{i}\cdot \gamma _{j}\psi ,\psi )\gamma
^{i}\wedge \gamma ^{j}  \tag{33}
\end{equation}

of the S-W equations. Such Bogomolny-type calculation is depended upon the

assumption,Eq.(33), playing crucial role in S-W formalism but, thus far, not

implied by the atomic physics formalism. This deficiency is corrected in the

Appendix E.

\ \ \ \ \ With these results established, from Eq.(31) it follows that in
the case when the

scalar curvature $R>0,$ the set of S-W equations, just defined, contain only
the

trivial solution: $\mathbf{A}=0,\psi =0.$ The identity 
\begin{equation}
\frac{1}{2}\Delta \left\vert \Phi \right\vert ^{2}=(\nabla _{A}^{+}\nabla
_{A}\Phi ,\Phi )-(\nabla _{A}\Phi ,\nabla _{A}\Phi )  \tag{34a}
\end{equation}

implies%
\begin{equation}
\frac{1}{2}\Delta \left\vert \Phi \right\vert ^{2}\leq (\nabla
_{A}^{+}\nabla _{A}\Phi ,\Phi ).  \tag{34b}
\end{equation}

Using Eq.(30) in this inequality, we obtain:%
\begin{equation}
\Delta \left\vert \psi \right\vert ^{2}\leq -\frac{R}{2}\left\vert \psi
\right\vert ^{2}-(F^{+}(A)\bar{\psi},\psi ).  \tag{35a}
\end{equation}

By comparing Eq.s(30) and (31) we can rewrite the last result as%
\begin{equation}
\Delta \left\vert \psi \right\vert ^{2}\leq -\frac{R}{2}\left\vert \psi
\right\vert ^{2}-\frac{1}{4}\left\vert \psi \right\vert ^{4}.  \tag{35b}
\end{equation}

This result can be integrated. If the boundary conditions are chosen
appropriately, such

an integration along with use of the Cauchy-Schwartz inequality and
normalization of $\psi $

produces:%
\begin{equation}
\frac{1}{2}\dint\limits_{M}\left\vert \psi \right\vert ^{4}dvol\leq
-\dint\limits_{M}R\left\vert \psi \right\vert ^{2}dvol\leq
\dint\limits_{M}R^{2}dvol.  \tag{36}
\end{equation}

Following Jost[$35$] and by using Eq.s(31)-(33) we obtain finally: 
\begin{equation}
\dint\limits_{M}\left\vert F(A)^{+}\right\vert ^{2}dvol\leq \dint\limits_{M}%
\frac{R^{2}}{4}dvol.  \tag{37}
\end{equation}

Up to extra factor of $\frac{1}{4}$in the r.h.s., this inequality coincides
with the inequality,

Eq.(3.5), obtained in the seminal paper by Witten[$56$]. In our case, the
obtained inequality

should be interpreted differently. The scalar curvature-the product of
electron-electron

interactions-controls existence or nonexistence of the $spin^{c}$ phase
since $F(A)^{+}$ can

be used only when such $spin^{c}$ structures are topologically \ permitted.
That is, the

Madelung anomalies typically cannot occur in atoms with low electron
content. This,

physically plausible, result can be considerably enhanced using the concept
of the

moduli space. Very much like in the Yang-Mills case, the S-W functional as
well as

the S-W equations should be \ invariant with respect to the gauge
transformations. In

the present case the gauge group $\mathcal{G}$ is made of maps $f$ - from M
to $S^{1}.$ Let $u\in \mathcal{G}$,

then the gauge transformations of S-W equations are described by (not to be
confused

with the scalar multiplication): 
\begin{equation}
(A,\psi )\longrightarrow (A+2u^{-1}du,u^{-1}\psi )  \tag{38}
\end{equation}

The moduli space $\mathcal{M}$ are formally defined as a quotient 
\begin{equation}
\mathcal{M}=solutions/\mathcal{G}  \tag{39}
\end{equation}

The most relevant for us is the case when $\mathcal{M=}0$. In such a case
the S-W equations

possess only finite number of localized solutions. \ This fact provides
justification for

the existence of finite number of Madelung anomalous solutions. It the
Appendix C

we outline different approach to this result.

Surprisingly, in the version of S-W theory considered by Witten$[56],$ the
case

$\mathcal{M=}0$ happens to be also the most interesting one. It is
associated with the fact

that the manifold M posses almost complex structure [$33$], e.g. read the
Lemma

on page 89. In physical language, this means that the (semi)classical limit
of

quantum mechanics on such manifolds are well defined because they admit

well defined classical trajectories. This conclusion was reached by Witten$%
[56]$

already in his first original paper on the subject.

\ \ \ \ \ Mathematically, the condition $\mathcal{M}=0$ occurs for manifold
M 's for

which \ 
\begin{equation}
c_{1}(L^{2})^{2}[\text{M}]=3\tau \lbrack \text{M}]+2\chi \lbrack \text{M}],%
\text{ }2c(L)=c(L^{2}).  \tag{40}
\end{equation}

Here $\tau \lbrack $M$]$ is the signature and $\chi \lbrack $M$]$ is the
Euler's characteristic of M, while $c_{1}(L)$ is

the 1st Chern class of the line bundle $L$ (connected with $S^{1}$ for the $%
spin^{c}$ manifolds

as explained in the \ Appendix D). Connection with\ the Atiyah-Singer index
theorem

can be seen directly by reading page 64 of Ref.[$33$]. It is associated with
the vanishing

of the 2nd Chern class $c_{2}$($S_{n}^{+}\otimes L)[$M$].$ Here $S_{n}^{+}$
is a part of the \textit{spinor bundle} defined in

the Appendix D. In the Appendix E 4. physically motivated explanation of

$spin^{c}$ manifolds is given. \bigskip It is associated with phenomenon of
superconductivity. Such

an explanation \ is plausible (makes sense) since the S-W theory is
reducible to the G-L

theory whose origins are in superconductivity [$50$].\medskip \medskip

\textbf{6. \ From the Madelung-exceptional atoms to Madelung-exceptional
solids}

\ \ \textbf{\ \ }

\ \ \ \ To our knowledge, in this work for the first time the
superconducting nature of the

Madelung-exceptional atoms is elucidated. Historically, however, study of

superconductivity at small scales has its beginnings in nuclear physics. It

was initiated immediately after the development of superconductivity theory
in metals

and alloys [$61$], [$66$]. \ Obviously, for the atomic nuclei as much as for
the hydrogen atom,

or the Madelung-exceptional atoms, there is no point to talk about the
macroscopic

evidence of superconductivity. The superconductivity for these systems
should be

understood in terms of Bogoliubov's quasiaverages introduced in subsection
1.2. This

means that breaking of U(1) gauge invariance associated with nonconservation

of Cooper pairs \ is mathematically reflected in emergence of \
quasiaverages [$13$],[$14$].

At the scales of atomic nuclei the superconductive properties should be
studied

spectroscopically since the U(1) invariance \ and its violation is related
to the

electromagnetic field.

\ \ \ \ \ \ \ \ \ By extending this direction of thought, P.W.Anderson in
1959 formulated the

following problem [$74$]. Suppose we have a metallic superconductor. Suppose
that

we can make a powder from it containing smaller and smaller grains. Then,
there will

be a grain size such that it will loose its superconducting properties.
Notice, though,

that such grains are expected to be larger in size than the atomic nuclei.
Nevertheless,

if in nuclear physics the concept of superconducting nuclei is firmly
established

spectroscopically, the same must be true for the superconducting grains. The
spectroscopy

of such granular materials was discussed in great detail in [$75$]. Since
the spectroscopy is

working for granular superconductors, it should be working as well for \ the
Madelung-

exceptional atoms. \ At the same time, when solids are made of such atoms,
the reversible

hydrogen absorption becomes indicative of Bogoliubov's quasiaverages. Then
it becomes

appropriate to talk about the superconductivity by applying \ the concept of
\ quasiaverages

to the reversible absorption. To do so requires the Madelung-exceptional
metal to be

placed into gaseous hydrogen environment.

\ \ \ \ Since nuclear excitations are, in fact, the excitations of
quark-gluon plasma, the nuclear

spectroscopy should seamlessly merge with the spectroscopy of hadrons and,
hence,

with excitations of the Yang-Mills fields. Such line of research was
initiated in our works

[$76$],[$77$]. This fact allows us \ to reduce the discussion in this
section to the minimum.

Also, it is fortunate that some of the methods we are about to discuss
recently have found

their place in chemistry [$78$].

\ \ \ In view of just made comments and to put things in the correct
perspective, we still

would \ like to make several remarks. \textsl{First}, Eq.s (4) and (25) are
manifestations of the

Weitzenbock formula%
\begin{equation}
\mathcal{D}^{2}\psi =0,\text{where }\mathcal{D}^{2}=\nabla ^{\ast }\nabla +%
\text{\textbf{K.}}  \tag{41}
\end{equation}

Here $\mathcal{D}^{2}$, defined in Eq.(E.12) , is the Hodge Laplacian. For
the $spin$ manifolds \textbf{K}= $\mathcal{R}^{S}$ \ 

implying that we are dealing with Eq.(4) while for the $spin^{c}$ manifolds 
\textbf{K}= $\mathcal{R}^{S}$+$\mathcal{F}^{S\text{ }},$

e.g. see $($Eq$.$E$.16)$ and \ we are dealing with Eq.(25). As $\exp $lained
in the Appendix

E.4., the superconductivity takes place on the $spin^{c}$ manifolds only. 
\textsl{Second}, the

Hodge Laplace, Eq.(41), is just the linearized analog of the respective
Hodge-like

Laplacian-type equation for the Yang-Mills fields [$79$]. Thus, the nuclear

superconductivity excitations are, in fact, also excitations of the
Yang-Mills fields.

\ Since our Ref.[$77$] is devoted entirely to study of the millennium
Yang-Mills gap

problem, it is hoped, that the present work might provide eventually its own

contribution to the gap problem.

\ \ \ Following [$76$],[$77$] as well as [$80$],[$81$] and using Eq.(E.20),
we begin with the

Hamiltonian, \^{H}=\^{H}$_{0}$ +\^{H}$_{V}$, and replace the matrix element V%
$_{\mathbf{kk}^{\prime }}^{(0)}$ by the constant $-g$.

The resulting Hamiltonian is given then by Eq.(5.42) of [$77$], i. e. 
\begin{equation}
\text{\^{H}}=\dsum\limits_{f}2\varepsilon _{f}\hat{N}_{f}-g\dsum\limits_{f}%
\dsum\limits_{f^{\prime }}\hat{b}_{f}^{+}\hat{b}_{f},  \tag{42}
\end{equation}

where \^{N}$_{f}=\frac{1}{2}(c_{f\ \ +}^{+}c_{f\ \ +}+c_{f\ \ -}^{+}c_{f\ \
-}),\hat{b}_{f}=c_{f\ \ -}c_{f\ \ +}.$ Here the operators $c_{f\ \ \sigma
}^{+}$

and $c_{f\ \ \sigma },$ $\sigma =\pm $ are obeying the usual anticommutation
relations for fermions:

\{$c_{f\ \ \sigma },c_{f^{\prime }\ \ \sigma ^{\prime }}^{+}\}=\delta
_{\sigma \sigma ^{\prime }}\delta _{ff^{\prime }}$ . Having these results
defined, it is convenient to

introduce the \textsl{seniority }operator $[80$] : $\hat{\nu}_{f}=c_{f\ \
+}^{+}c_{f\ \ +}-c_{f\ \ -}^{+}c_{f\ \ -}.$It is taking care of

the number of unpaired fermions at each level $f$. By construction, [\^{H}, 
\^{N}$_{f}]$

$=[$\^{H}, $\hat{\nu}_{f}]=0$. These commutators permit us to make a
subdivision \^{H}=\^{H}$_{1}$

+\^{H}$_{2}$ and to count configurations beginning with the situation when

$g=0$ since the eigenvalues $\nu _{f}$ of the seniority operator $\hat{\nu}%
_{f}$ (0 and $\sigma )$

remain unaffected by $g$. In such a case let \^{H}$_{1}$ describe the states
without

Cooper pairs, that is it describes the Hilbert space sector for which $\nu
_{f}$ =$\sigma .$

Accordingly, \^{H}$_{2}$ is to be associated with the sector for which $\nu
_{f}$ =$0.$ Such a

subdivision produces a remarkable and unexpected result: the matrix elements

of \ \^{H}$_{2}$ are calculated with help of the \textbf{bosonic-type}
commutation relations.

These are%
\begin{equation}
\lbrack \hat{b}_{f},\hat{N}_{f^{\prime }}]=\delta _{ff^{\prime }}b_{f}\text{
\ and [}\hat{b}_{f},\hat{b}_{f}^{+}]=\delta _{ff^{\prime }}(1-2\hat{N}%
_{f^{\prime }})  \tag{43}
\end{equation}

Even though these are bosonic commutators, but they are nontraditional ones.

In the traditional case we would have [$\hat{b}_{f},\hat{b}_{f}^{+}]=\delta
_{ff^{\prime }}.$ To by pass the emerging

difficulty is equivalent to solving the eigenvalue problem for the
Hamiltonian,

Eq.(42). This was done in [$80$], but more elegantly in [$81$]. The problem
was

reduced \ to finding the spectrum of the Richardson-Gaudin one dimensional

spin chain. Its excitation spectrum resemble that for the nonideal Bose gas.

Before writing down this spectrum of H, we define $\Omega _{n}$ as the pair
degeneracy

of the level $n$, that is $\Omega _{n}$ is the number of values of $f$ \ for
which $\varepsilon _{f}=\varepsilon _{n}$.Omitting

all the details given in [$76$],[$77$], [$80$],[$81$], we introduce the

function $F(E)$ via%
\begin{equation}
F(E)=\dsum\limits_{n}\Omega _{n}(2\varepsilon _{n}-E)^{-1}  \tag{44}
\end{equation}

so that the spectrum for just one Cooper pair is obtained graphically using
the

equation $F(E)=g^{-1}.$ This equation was initially obtained by Cooper [$82$%
]. It

paved a way for development of the BCS theory of superconductivity. To extend

this result for many Cooper pairs Richardson assumed that the wave function
for H

is made of a symmetrized product of $N$ Cooper pairs wave functions so that
the

total energy of $N$ pairs is the sum of the respective energies of $N$
Cooper pairs.

This assumption allows us to write the\ spectrum as 
\begin{equation}
F(E_{p_{i}})=g_{i}^{-1},\text{ }g_{i}=g[1+2g\dsum\limits_{j\neq
i}^{N}(E_{p_{j}}-E_{p_{i}})^{-1}]^{-1};i=1,...,N.  \tag{45}
\end{equation}

Presented results serve only to introduce our readers to more complicated
problems

such as a) crossover: from the atomic limit-to the bulk metal and, b)
effects of finite

temperatures. The crossover problem (even including the temperature effects)
was

discussed in detail in section 5 of [$75$]. \ Since this review was
published in 2001, we

decided to provide up to date (2020-2021) results. In [$83$] the results of
Richardson

and Sherman [$80$] were elaborated further. In [$84$] and [$85$] the results
of [$75$] were

significantly elaborated. In [$86$] the results on nanoclusters of high
temperature

superconductors were reported. \bigskip

\textbf{7. Summary and discussion\bigskip }

Study of the high temperature in superconductors [$5$] cannot progress
without discoveries

of new guiding principles. The purpose of this paper is to supply several
new guiding

principles. These are based on theoretical explanations of several empirical
observations.

These are: a) the majority of Madelung -exceptional elements yield the
highest T$_{c}^{\prime }s$ to

date, b) the Madelung exceptionality is linked with the property of
reversible hydrogen

absorption yielding exceptionally high concentration hydrides of these
elements [$10$] [$11$],

[$87$], pages79-84. Among all these Madelung-exceptional elements. The
Madelung-

exceptional elements were in use till now without emphasis on their Madelung

exceptionality. Empirically, this exceptionality was noticed due to the
unusual property-

the reversible hydrogen absorption. In this regard the most notable example
is palladium.

The electronic structure of palladium (Appendix F) makes it a benchmark
object of study.

From this angle it is not too surprising that it was Pd \ which was used
initially in the

cold fusion experiments \ https://en.wikipedia.org/wiki/Cold\_fusion .

The empirical ability to absorb large amount of hydrogen singed out
Madelung-exceptional

metals as likely candidates of high T$_{c}$ superconductors. The fact that
the hydrogen

absorption is reversible (under the appropriate experimental conditions $[88$%
]) caused us

to use Bogoliubov's method of quasiaverages [$13$],[$14$] associated, in the
present case

with Cooper pairs nonconservation. This property is characteristic of
superconductivity.

As we argued in section 6, the phenomenon of superconductivity exist on many
scales

[$89$]. When looking at the Madelung -exceptional elements one should not
anticipate all

of them to be superconductors in traditional sense without hydrogen
environment. \ But

once such environment is provided (under appropriate pressure-temperature
conditions)

they all are becoming superconductors in traditional sense. This is indeed
the case, for

example, for Pd. Pd \ is not a superconductor \ but in the presence of small
amount of

gaseous hydrogen it becomes superconducting \ under usual ambient conditions
(e.g. read

the subsection 1.2.). Since the superconductivity is observed at scales
ranging from

atomic nuclei to the scales of neutron stars [$89$], it is only natural to
search for

superconductivity at the atomic scales. It was done in this paper. Here it
is demonstrated

analytically (by non trivially solving the quantum many-body problem and
invoking some

results from S-W theory) that only the Madedulng-exceptional atoms possess
the

superconducting property. It might be detected spectroscopically eventually.
By solving

exactly\ the quantum many -body problem at the level of a single atom, we
briefly

sketched ways of extending the obtained results to the atomic clusters by
relying on

methods developed for the superconducting clusters. Since the size of
clusters is an

adjustable parameter in these calculations, the problem of crossover- from
the atomic

scales to scales of bulk metals- was briefly outlined as well. In this
regard, Ref.[$90$],

might serve as an excellent point of departure for further studies.

\bigskip

\bigskip \textbf{Acknowledgement \bigskip }

The author greatly benefited from countless conversations with Dr.Jack
Douglas (NIST)

as well as from very helpful comments and critique of the original version
of the

manuscript by the anonymous referee.

\bigskip

\textbf{APPENDIX A\bigskip . Mapping the Dirac equation into Schr\"{o}%
dinger-like }

Following Wong and Yeh $[39],$ we shall employ the system of units in which $%
c=1$,

$\hbar =1.$ Then, taking into account that in discussing Eq.s(16) and (17)
we introduced

the factor $\mathcal{K}$ in the nonrelativistic case and $\Gamma $ in the
relativistic, we argued that the

combination $\mathcal{K}(\mathcal{K}+1)=l(\kappa )(l(\kappa )+1)=l(l+1)$ in
the nonrelativistic case and

$\Gamma (\Gamma +1)=l(\gamma \kappa )(l(\gamma \kappa )+1)$ in the
relativistic. Therefore Eq.(17) acquires the form 
\begin{equation}
\lbrack \frac{1}{r^{2}}\frac{d}{dr}r^{2}\frac{d}{dr}-\frac{l(\gamma \kappa
)(l(\gamma \kappa )+1)}{r^{2}}+\frac{2ZEe^{2}}{r}+E^{2}-m^{2}]R_{N,l(\gamma
\kappa )}(r)=0.  \tag{A.1}
\end{equation}

The nonrelativistic $l$ $\ $is replaced now by the relativistic $\kappa =\pm
(j+\frac{1}{2})$ and

$\gamma \kappa =\pm \lbrack \kappa -\left( Ze^{2}\right) ^{2}],$ see
Eq.(19). In the case of discrete spectrum $m^{2}-E^{2}>0.$

Therefore, it is \ convenient to introduce the new variables as follows:

$\mu =[m^{2}-E^{2}]^{\frac{1}{2}},\rho =2\mu r,\mathfrak{\omega =}%
4Ze^{2}E/\mu .$ In terms of these variables Eq.(A.1)

acquires standard form of the radial equation for hydrogen atom%
\begin{equation}
\lbrack \frac{1}{\rho ^{2}}\frac{d}{d\rho }\rho ^{2}\frac{d}{d\rho }-\frac{%
l(\kappa )(l(\kappa )+1)}{\rho ^{2}}+\frac{\omega }{4\rho }-\frac{1}{4}%
]R_{N,l(\gamma \kappa )}(\rho )=0  \tag{A.2}
\end{equation}

e.g. see,$[48],$ Eq.(16.7).\bigskip\ This transformation allows us to apply
unchanged

the same methodology as was developed in $[3]$ for proving the standard

Madelung rule. \bigskip

\bigskip

\textbf{APPENDIX B.\ Mapping of the Coulombic potential problem into}

\ \ \ \ \ \ \ \ \ \ \ \ \ \ \ \ \ \ \ \ \ \ \ \ \ 

\ \ \ \ \ \ \ \ \ \ \ \ \ \ \ \ \ \ \ \ \ \ \ \ \ \textbf{the fish-eye
problem\bigskip . Emergence of conformal}

\textbf{\ \ \ \ \ \ \ \ \ \ \ \ \ \ \ \ \ \ \ \ \ \ \ \ invariance }

\ \ 

As demonstrated by Schr\"{o}dinger in his 1st paper on quantum mechanics

the standard Schr\"{o}dinger equation%
\begin{equation}
\left( \frac{\partial ^{2}}{\partial x^{2}}+\frac{\partial ^{2}}{\partial
y^{2}}+\frac{\partial ^{2}}{\partial z^{2}}\right) \varphi =\frac{2m}{\hbar
^{2}}(E-V)\varphi  \tag{B.1}
\end{equation}

is obtainable variationally from the Hamilton-Jacobi equation$[34]$

\begin{equation}
\psi _{x}^{2}+\psi _{y}^{2}+\psi _{z}^{2}=2m(E-V),  \tag{B.2}
\end{equation}

where $\psi \rightleftarrows \hbar \ln \varphi $. Next, following Luneburg[$%
22$] we use the canonical change of variables:

$\xi =\psi _{x},\eta =\psi _{y},\zeta =\psi _{z};x=\omega _{\xi },y=\omega
_{\eta },z=\omega _{\zeta }$ subject to the condition

$\psi +\omega =x\xi +y\eta +z\zeta $ in Eq.(B.2).\ When $V$ is the attractive

Coulombic potential and such transformations are applied to Eq.(B.2) the
result is 
\begin{equation}
\frac{1}{2}\left( \frac{\partial ^{2}}{\partial x^{2}}+\frac{\partial ^{2}}{%
\partial y^{2}}+\frac{\partial ^{2}}{\partial z^{2}}\right) \varphi +\beta
_{n}\left( \frac{1}{1+x^{2}+y^{2}+z^{2}}\right) ^{2}\varphi =0,  \tag{B.3}
\end{equation}

where $\beta _{n}=\left( \frac{Ze}{E_{n}}\right) ^{2}.$ Just described chain
of transformations is converting the eigenvalue

problem, for Eq.(B.1), into the Sturmian problem, for Eq.(B.3). The Coulombic

potential $V_{C}=\frac{const}{r}$ in Eq.(B.1) is converted into the
Maxwell's fish-eye potential

$V_{F}=\frac{const^{\prime }}{1+\left( r/a\right) ^{2}}%
,r^{2}=x^{2}+y^{2}+z^{2}.$ Here $a$ is a constant In Eq.(B.3) \ we select $%
a=1$.

The conversion into the Sturmian problem\ is having an additional advantage.
It

is converting Eq.(3) into conformally invariant Eq.(4) [$3$]. Uses of
conformal

transformations then allows us to recreate \textbf{exactly} the effects \ of
many-body

electron-electron interactions (at this stage of our study, formally,
without explicit

account of the spin-spin interactions). These are accommodated into the
formalism

with help of the results of section 4 and Appendix E. Applications of
conformal

transformations to Eq.(B.3) converts Maxwell's fish-eye potential $V_{F}$
into its

conformally deformed form:%
\begin{equation}
V(r)\equiv V(x,y,z)=-\left( \frac{a}{r}\right) ^{2}\left[ \frac{n_{0}}{%
\left( r/a\right) ^{-\gamma }+\left( r/a\right) ^{\gamma }}\right] ^{2}. 
\tag{B.4}
\end{equation}

This results in replacement: $V_{eff}=V_{F}$ in\ Eq.(4) by the $V(r),$
Eq.(B.4). Details are given

in [$3$]. Application of scaling analysis-the simplest of conformal
transformations-indicates

that only two exponents $\gamma $ are permissible. Using $\gamma =1$
recreates back the Coulombic

fish-eye potential for the hydrogen atom. Using $\gamma =1/2$ recreates the
multielectron

effects for any multielectron atom. Furthermore, using Eq.(B.4) with $\gamma
=1/2$ \ 

numerically produces \textbf{exactly} \textbf{the same} results as those
known for the Hatree-Fock

potential. The value of the constant $\beta _{n}$ is not affected by uses of
conformal

transformations. The additional bonus of using $V(r),$ Eq.(B.4), is coming
from the

very nontrivial exact conversion of the $V(r)$ into potential obtained by
Perlick [$23$]

in his studies of the \ generalized Bertrand theorem.

\bigskip

\textbf{APPENDIX C}. \textbf{Calculation of the hydrogen and the
Madelung-regular}

\ \ \ \ \ \ \ \ \ \ \ \ \ \ \ \ \ \ \ \ \ \ \ \ \textbf{\ atomic spectra. \
Alternative explanation of the discretness}

\ \ \ \ \ \ \ \ \ \ \ \ \ \ \ \ \ \ \ \ \ \ \ \ \ \textbf{of moduli space\
\bigskip }

a) \textsl{Calculation of the spectrum}. To treat the accidental degeneracy
in the

spectrum of hydrogen atom Fock developed entirely new method of

solving the spectral problem for hydrogen atom$[34]$ by considering

the solution of this problem on $S^{3}$. Since Eq.(B 3) is not an eigenvalue

but the Sturmian problem, we cannot apply the Fock method as such.

However, we do apply his idea of replacing the treatment of Eq.(B 3)

in \textbf{R}$^{3}$ by the treatment on $S^{3}$ in accord with the results
of subsection 3.3$.$

By lifting Eq.(B3) to $S^{3}$ it is converted to Eq.$(11a)$. In Eq.(11a) we
have to

present $Y_{nlm}(\alpha ,\theta ,\phi )$ as $\Psi _{nl}(\alpha
)Y_{lm}(\theta ,\varphi )$ so that Eq.(11a) acquires the

form:%
\begin{equation}
\lbrack \frac{l(l+1)}{\sin ^{2}\alpha }-\frac{\partial ^{2}}{\partial \alpha
^{2}}-2\cot \alpha \frac{\partial }{\partial \alpha }]\Psi _{nl}(\alpha
)=I_{nl}\Psi _{nl}(\alpha ).  \tag{C 1}
\end{equation}

Here $I_{nl}=-\frac{\left( Ze\right) ^{2}}{2\left\vert E_{n}\right\vert }$
in the nonrelativistic case. In the relativistic case

we have to make a replacement $l\rightarrow l(\kappa )$ and \ to write $%
I_{nl}=\left( \dfrac{\omega }{4}\right) ^{2}.$

All details are given in the Appendix A and must be performed with account
that

$E$ is describing bound states. Next, we write $x=cos\alpha $ and, by
rewriting $\Psi _{nl}(\alpha )$

in terms of such variable and representing it in the form $\Psi _{nl}(\alpha
)=(1-x^{2})^{\frac{l}{2}}F_{nl}(x),$

Eq.(C1) is converted into equation%
\begin{equation}
(1-x^{2})\frac{d^{2}}{dx^{2}}F_{nl}(x)-(2x+1)x\frac{d}{dx}%
F_{nl}(x)+[I_{nl}-(l(l+2)]F_{nl}(x)=0.  \tag{C 2}
\end{equation}

Eq.(C 2) is the equation for the Gegenbauer polynomials. Using this fact

we obtain, after some calculation: $I_{nl}=(n+l+1)^{2}-1\equiv \tilde{n}%
^{2}-1.$

Let now $\tilde{n}=2F+1$. Then, $\tilde{n}^{2}-1=4F(F+1).$ Using this
information,

consider, instead of Eq.(11a) (that is Eq.(C.1)) the equation $\mathcal{L}%
^{2}Y_{nlm}=\left( I_{nl}-E\right) Y_{nlm}$

in which $E$ is the fixed parameter. \ The necessity of doing this is
explained in the

section 4 and Appendix F of [$3$]. To determine the value of this parameter
we

analyze the equation $I_{nl}-E=4F(F+1).$By selecting $-E=-1$ we obtain:

$I_{nl}=(2F+1)^{2}$ implying%
\begin{equation}
\frac{-\left( Ze\right) ^{2}}{2\left\vert E_{nl}\right\vert }=\tilde{n}^{2}%
\text{ or, }E_{nl}=\frac{-\left( Ze\right) ^{2}}{2\tilde{n}^{2}}=\frac{%
-\left( Ze\right) ^{2}}{2(n+l+1)^{2}}\text{ \ \ \ \ (Schr\"{o}dinger
spectrum).}  \tag{C 3}
\end{equation}

Here $n=n_{r}$ in the standard quantum mechanical notations. For

the Dirac case we obtain as well:%
\begin{equation}
\left( \dfrac{\omega }{4}\right) ^{2}=(n_{r}\text{ +}l(\gamma \kappa )+1)^{2}%
\text{ \ (Dirac spectrum)},  \tag{C 4}
\end{equation}

where $\omega $ is defined in the Appendix A.

The result, Eq.(C.4), coincides with Eq.(3.26) of Ref.[$39$]. By restoring
back $c$, $\hbar $ and,

hence, the $\alpha ,$and using Eq.(C 4) we reobtain back the known Dirac
spectrum.\bigskip \bigskip \bigskip

The Madelung-regular spectrum emerges as solution of Eq.(11b). In view of
the fact

that Eq.(11b) emerges as modification of Eq.(11a) caused by change: from $%
\gamma =1$ to

$\gamma =1/2$ in Eq.(B.4). Eq.(C.1) changes accordingly also. This leads to
some changes in

Eq.(C.2) while keeping $I_{nl}=-\frac{\left( Ze\right) ^{2}}{2\left\vert
E_{n}\right\vert }$ unchanged. Due to changes in Eq.(C.2), the

spectrum, that is the Eq.(C.3), also changes resulting in Eq.(27) of the
main text.

Details are given in [$3$].

b) \textsl{Discretness of the moduli space}. \medskip As it follows from
results of section 5, we need to

provide an evidence that, upon relativization, not all atoms become the
Madelung-

exceptional. We would like to achieve this by using the inequality,Eq.(36).
In this

inequality we shall choose $M=S^{3}.$ On $S^{3}$ we shall use properly
normalized spherical

eigenfunctions \ Y$_{n+l,,l}m(\alpha ,\theta ,\varphi )$ defined in
Eq.(11b), see also [18]. \ Account of

relativistic effects leads to the replacement of quantum number $l$ by the $%
l(\gamma \kappa ),$

defined in Eq.(A.1). Also, instead of $I_{nl}=-\frac{\left( Ze\right) ^{2}}{%
2\left\vert E_{n}\right\vert }=\tilde{n}^{2}$ , in view of Eq.(C.3), we

have to use ($I_{nl})^{2}$ (for $R^{2}),$where\ now\ \ $I_{nl}=(n_{r}$ +$%
l(\gamma \kappa )+1)^{2}$, in view of Eq.(C.4).

Since the area of $S^{3}$ is known constant, the inequality, Eq.(3.6), is
regulated by the

charge $Z$ of the atomic nucleus. When the inequality becomes an equality in
Eq.(36),

it provides a complicated equation for Z whose acceptable solutions should
be only

in terms of the integer nonnegative Z's. It is clear then, that there could
be \ only a

countable number of Z's or no Z's at all. The last case brings us back to
the Madelung

-regular case which does not require such an inequality. This is so because
the

resulting from Eq.(36) equality will not contain the parameter Z at all and,

therefore, neither the inequality, Eq.(36), nor the equality originated from
Eq.(36),

make no physical and, even mathematical, sense when Z is absent. That is
relativistic

effects are completely ignored.

\bigskip

\textbf{APPENDIX D. } \textbf{Spin structures.} \textbf{Group-theoretical
and topological}

\ \ \ \ \ \ \ \ \ \ \ \ \ \ \ \ \ \ \ \ \ \ \ \ \ \ \ \textbf{%
aspects\bigskip }

In subsection 4.1. the anticommutator relation defining Clifford algebra was
presented.

In this and next appendix \ we develop quantum many-body formalism using
Clifford

algebras. We begin with\medskip \medskip

\textbf{D}.\textbf{1. Vector and spinor representations of Clifford
algebras\bigskip }

Let $V$ be some vector space of dimension $n$ over \textbf{R} and $g$ being
some non degenerate

bilinear form on $V$. The Clifford algebra $Cl(V,g)$ is an associative
algebra \ with unit

defined by 
\begin{equation}
Cl(V,g)=\frac{T(V)}{I(V,g)},  \tag{D.1}
\end{equation}

where $T(V)$ is the tensor algebra and $I(V,g)$ is the ideal created by $%
x\otimes x+g(x,x)1,\forall x\in V.$

If we define a map $x\rightarrow c(x)$ such that $x\otimes
x+g(x,x)1\rightarrow c(x)\otimes c(x)+g(c(x),c(x))1,$

then, there is a unique algebra homeomorphism: $Cl(V,g(x))\rightarrow
Cl(c(V),g(c(x)).$ In such

a way in section 4.1. we replaced $\gamma ^{a}\gamma ^{b}+\gamma ^{b}\gamma
^{a}=2\eta ^{ab}$ by $\gamma ^{\mu }\gamma ^{\nu }+\gamma ^{\nu }\gamma
^{\mu }=2g^{\mu \nu }$ so that the

c-map is $\gamma ^{\mu }=e_{a}^{\mu }\gamma ^{a}.$ Clearly, other options
for c-maps are possible as well.

Let now ($e_{1},...,e_{n})$ be a $g$-orthonormal basis of $V$, then 
\begin{equation}
\{e_{0}:=1,e_{k}:=e_{i_{1}}\cdot \cdot \cdot e_{i_{k}}\mid 1\leq i_{1}<\cdot
\cdot \cdot <i_{k}\leq n;\text{ \ }0\leq k\leq n\}  \tag{D. 2}
\end{equation}

is the basis of $Cl(V,g)$ with dimension $\dim Cl(V,g)=2^{n}.$ There is a
canonical

isomorphism of vector spaces (as algebras) between the exterior algebra and
the Clifford

algebra $\Lambda ^{\ast }V\longrightarrow Cl(V,g),$ that is%
\begin{equation}
e_{i_{1}}\wedge \cdot \cdot \cdot \wedge e_{i_{k}}\rightarrow e_{i_{1}}\cdot
\cdot \cdot e_{i_{k}}.  \tag{D.3}
\end{equation}

This fact is compatible with the observation that the ideal $I(V,g=0)$ in
Eq.(D.1) converts

the Clifford algebra into the Grassmann algebra. Thus, the Clifford algebra
is the

deformation of the Grassmann algebra. The canonical isomorphism, Eq.(D.3),
makes

it possible (and mathematically even necessary) to replace all Grassmann
algebra

results in physics literature by those involving the Clifford algebra. More
details

are presented in the Appendix E below.

\ \ \ \ The above isomorphism does not depend upon the choice of the basis
of $V$. The anti

automorphism $t$ of $Cl(V,g)$ is defined as ($e_{i_{1}}\cdot \cdot \cdot
e_{i_{k}})^{t}=e_{i_{k}}\cdot \cdot \cdot e_{i_{1}}(=(-1)^{\frac{k(k-1)}{2}%
}e_{i_{1}}\cdot \cdot \cdot e_{i_{k}}).$

Thus, $e_{i_{1}}\cdot \cdot \cdot e_{i_{k}}$($e_{i_{1}}\cdot \cdot \cdot
e_{i_{k}})^{t}=1($if $k$ is even) and $=-1$ (if $k$ is odd). Using these

definitions we are in the position to define $\ $the $Pin(V)$ and $Spin(V)$
groups. Specifically,

$Pin(V)$ is the group of elements \ $a\in $ $Cl(V,g)$ such that%
\begin{equation}
Pin(V):\{a=e_{1}\cdot \cdot \cdot e_{k}\mid g(e_{i},e_{i})=1,\forall i=1\div
k\},  \tag{D.4a}
\end{equation}

while 
\begin{equation}
Spin(V):\{a=e_{1}\cdot \cdot \cdot e_{2k}\mid g(e_{i},e_{i})=1,\forall
i=1\div 2k\}.  \tag{D.4b}
\end{equation}

By design, for $Spin(V)$ $aa^{t}=1.$ Let $\rho (a)v=ava^{t}$, $v\in V$, then
one can construct a

surjective homeomorphism $\rho \left( Pin(V)\right) \rightarrow O(V)$ while
use of $Spin(V)$ results in

$\rho \left( Spin(V)\right) \rightarrow SO(V).$ By employing these
homeomorphisms it can be demonstrated,

that the group $Spin(V)$ is \textbf{universal double cover} of the group $%
SO(V).$ Its kernel is

determined by the equation $\rho (a)v=v,$ $\ \forall v\in V.$

Since in this case $aa^{t}=1,$we can rewrite the same equation as $av=va$
producing

the kernel ( fixed point): $a=\pm 1.$The obtained result allows us to make
further step by

defining the $Spin^{c}(V)$ group. To do so, requires some preparations.

\textbf{First}, we have to define the complexified Clifford algebra $Cl^{%
\mathbf{C}}(V)=Cl(V,g)\otimes _{\mathbf{R}}\mathbf{C}$.

\textbf{Second}, we have to define the \textit{chirality operator} $\Gamma $
via%
\begin{equation}
\Gamma =i^{m}e_{1}...e_{n}\in Cl^{\mathbf{C}}(V)  \tag{D.5}
\end{equation}

so that $m=n/2$ for $n$ even and $m=\left( n+1\right) $/2 for $n$ odd.
Evidently, $\Gamma ^{2}=1,\Gamma v=v\Gamma $ for

even $n$ and $\Gamma v=-v\Gamma $ for odd $n.$ That is $\Gamma $ is operator
of involution. This induces a

decomposition of $Cl^{\mathbf{C}}(V)$ into $Cl^{\mathbf{C}}(V)^{\pm }$
parts. The decomposition is a bit involved.

\textbf{Third}, in $V\otimes \mathbf{C}$ space we must:

$a$) introduce a subspace $W$ made of vectors 
\begin{equation}
\eta _{j}=\frac{1}{\sqrt{2}}(e_{2j-1}-e_{2j}),\text{ }j=1,...m,  \tag{D.6}
\end{equation}

$b$) to extend (to \textbf{C)} the Hermitian scalar product in such a way
that 
\begin{equation}
<\eta _{i},\eta _{j}>_{\mathbf{C}}=0\text{ }\forall j.  \tag{D.7}
\end{equation}

This is done with the purpose of introducing the dual space $\bar{W}$ via%
\begin{equation}
\bar{\eta}_{j}=\frac{1}{\sqrt{2}}(e_{2j-1}+e_{2j}),j=1,...m,  \tag{D.8}
\end{equation}

so that, instead of Eq.(D.7), we obtain: 
\begin{equation}
<\eta _{j},\bar{\eta}_{j}>_{\mathbf{C}}\equiv \left\Vert \eta
_{j}\right\Vert \text{ }\forall j.  \tag{D.9}
\end{equation}

Clearly, Eq.(D.9)\textsl{\ has a quantum mechanical meaning} to be amplified
below.\medskip

\textbf{Definition D.1}. The spinor space $S_{n}$ is defined as an exterior
algebra $\wedge W$ of $V$

(whose dimension is $n)$.\bigskip

Let $v=w+\bar{w}.$ Then, $\forall s\in S_{n}=\wedge W$ the endomorphism End$%
_{\mathbf{C}}(S_{n})$ denoted as $\rho (w)$

is defined as 
\begin{eqnarray}
\rho (w)s &:&=\sqrt{2}\varepsilon (w)s,  \TCItag{D.10a} \\
\rho (\bar{w})s &:&=-\sqrt{2}i(\bar{w})s.  \TCItag{D.10b}
\end{eqnarray}

Here $s=\eta _{j_{1}}\wedge \cdot \cdot \cdot \wedge \eta _{j_{k}};$ $1\leq
j_{1}<\cdot \cdot \cdot <j_{k}\leq m;\varepsilon (\eta _{j})s=\eta
_{j}\wedge \eta _{j_{1}}\wedge \cdot \cdot \cdot \wedge \eta _{j_{k}};$%
\begin{equation}
i(\bar{\eta}_{j})s:=\left\{ 
\begin{array}{c}
0\text{ if }j\neq \{j_{1},...,j_{k}\} \\ 
\left( -1\right) ^{l-1}\bar{\eta}_{j}\wedge \cdot \cdot \cdot \wedge \left( 
\bar{\eta}\right) _{l}^{0}\wedge \cdot \cdot \cdot \wedge \eta _{j_{k}}\text{
if }j=j_{l}%
\end{array}%
\right.  \tag{D.11}
\end{equation}

Here $\left( \bar{\eta}\right) _{l}^{0}$ denotes the term absent in the
exterior product. Thus introduced operator $\rho $

possesses the group representation property: $\rho (vw)=\rho (v)\rho (w).$
The chirality operator $\Gamma $,\ 

when rewritten accordingly in terms of $\eta $ and $\bar{\eta}$, allows us
to decompose $S_{n}$ for even $n$ as

follows 
\begin{equation}
S_{n}=S_{n}^{+}\oplus S_{n}^{-}.  \tag{D.12}
\end{equation}

Here $S_{n}^{+}$ and $S_{n}^{-}$ are eigenfunctions (half spinors) of the
operator $\Gamma $ whose eigenvalues

are $\pm 1.$

\bigskip

\textbf{Definition D.2.} The representation $\rho $ of $Spin(V)$ given by
Eq.s(D.10a) and (D.10b) \ on

the spinor space $S_{n}$ is called \textit{spinor representation.}The same,
but on $S_{n}^{+}$ and $S_{n}^{-}$ is called

\textit{half spinor representation. \medskip }

\textsl{Spinor representation is an unitary presentation }[$33$]\textsl{. It
preserves the Hermitian product.}

\textsl{Therefore it is ideally suitable for the quantum mechanical
calculations.}

To extend these results for odd $n$, it is helpful to know $[35]$ that : a)
for dim$V=2n,$

$Cl^{\mathbf{C}}(V)\simeq \mathbf{C}^{2^{n}\times 2^{n}},$ b) for dim$%
V=2n+1,Cl^{\mathbf{C}}(V)\simeq \mathbf{C}^{2^{n}\times 2^{n}}\oplus \mathbf{%
C}^{2^{n}\times 2^{n}}.$

With this information, the odd dimensional space $V$ does not create
additional

problems. The spinor and half spinor representations admit unique extension
to

$Spin^{c}(V).\medskip $

\textbf{Definition \ D.3}. A group $Spin^{c}(V)$ is a subgroup of
multiplicative group of units

(that is of elements having an inverse) of $Cl^{\mathbf{C}}(V).$ It is
generated as a surjective mapping:

$Spin(V)\times S^{1}\longrightarrow Spin^{c}(V),$ where $S^{1}$ is the unit
circle in $\mathbf{C.}$ If $a\in Spin(V)$ and $z\in S^{1}$

then, the kernel of this mapping is $az=1$ is implying: $a=z^{-1}\in
Spin(V)\cap S^{1}.\bigskip $

\textbf{Definition D.4}. $Spin^{c}(V)$ is isomorphic \ to $Spin(V)\times _{%
\mathbf{Z}_{2}}S^{1},$ where the \textbf{Z}$_{2}$ action

identifies $(a,z)$ with $(-a,-z).$ $Spin^{c}(V)$ yields a nontrivial double
covering

$Spin^{c}(V)\longrightarrow $ $SO(V)\times S^{1}.\medskip $

The physical meaning of $Spin^{c}(V)$ was never discussed in mathematical
literature.

It is explained below, in the subsection E. 4.

\medskip\ \medskip

\textbf{D.2.} \textbf{\ Spinor and Clifford bundles \medskip \bigskip }

If $TM$ is the tangent bundle of $M$, the Riemannian metric on $M$ \ reduces
the structure

group of $TM$ to $SO(n),n=\dim M.$ This fact allows us to design the \textit{%
associated} \ 

\textit{principal} bundle $P$ over $M$ with fiber $SO(n)$. Such an
associated bundle is called

\textsl{Clifford bundle }(see below). In general relativity such a bundle is

known as \textit{frame bundle}[$31$]\textit{.} Uses of vierbeins $e_{\mu
}^{a}(x)$ in subsection 4.1. reflect

just this fact. The spinorial analysis elevates this concept one level above
just

described. Specifically,\medskip\ it begins with the following\bigskip

\textbf{Definition D.5.} A \textit{spin structure} on $M$ is synonymous to
designing of the principal

bundle $\tilde{P}$ over $M$ \ with the fiber $Spin(n)$ (universal double
cover of $SO(n))$ for

which the quotient of each fiber by the center $\pm 1$ is isomorphic to the
frame bundle

just defined.\medskip

\textbf{Definition D.6. }A Riemannian manifold with a fixed spin structure
is called

\textit{spin manifold}. \medskip

Since \ the fiber $Spin(n)$ operates \ on the spinor space $S_{n},$
Eq.(D.12), and, for even $n,$

also on the half spinor spaces $S_{n}^{+}$ and $S_{n}^{-},$ it becomes
possible to talk about the spinor

bundle in this context.\medskip

\textbf{Definition D.7}. The \textit{Spinor bundle} $\mathfrak{S}_{n}$ is
defined as 
\begin{equation}
\mathfrak{S}_{n}=\tilde{P}\times _{Spin(n)}S_{n}  \tag{D.13}
\end{equation}

This definition is to be contrasted with the definition of \textit{Clifford
bundle}.\medskip

\textbf{Definition D.6. }Bundles $Cl(P)$ and $Cl^{\mathbf{C}}(P)$ defined
below%
\begin{eqnarray}
Cl(P) &=&P\times _{SO(n)}Cl(V),  \TCItag{D.14a} \\
Cl^{\mathbf{C}}(P) &=&P\times _{SO(n)}Cl^{\mathbf{C}}(V)  \TCItag{D.14b}
\end{eqnarray}

are called $Clifford$ $bundles$.\bigskip

From the definition of Clifford bundles it follows that the creation of such
bundles do

not require $\mathit{spin}$ or $spin^{c}$ structures. However, they can
exist on such structures as well.

This will be further studied in the next appendix in the context of Dirac
operators.

The fundamental issue\bigskip\ is: \textsl{If there a connection between the
Clifford and Spinor bundles}

, \textsl{what physics such a connections describes}?

To answer this question requires introduction of many nontrivial facts to be
described

below.\medskip \medskip

\textbf{APPENDIX E. Dirac operators on Clifford\ and\ spinor bundles\bigskip 
}

The purpose of this appendix is to demonstrate that the Eq.(25) obtained by
Schr\"{o}dinger

accounts for all quantum many-body effects for the atomic multielectron
system.

At the present, the relativistic many-body effects are treated with help of
the

relativistically extended Hartree-Fock variational methods $[24]$. In the

nonrelativistic limit the Hartree-Fock calculations end up with the
eigenvalue Eq.(3).

It does not obey the superposition principle though. This happens to be a
fundamental

problem for development \ quantum mechanics of many-body systems as
explained in

detail in the book by Tomonaga[$57$]. He calls the equations like Eq.(3) 
\textsl{De Broglie-type}.

He argues that the formalism of second quantization, essential for
development of

quantum field theory, is applicable only to the Sch\"{o}dinger-type
equations for which the

superposition principle holds. This is also explained in another of his
books[$58$], page108.

Ignoring the supersposition principle makes the underlying equations
formally purely

classical. That is in such equations the Plank constant $\hbar $ can be
eliminated by the

appropriate changes of variables and rescaling. This paradoxical situation
is explained

in detail in our work[$34$]. To our knowledge, in physics literature

the second quantization method is used in many-fermion theory with or
without account

of the superposition principle [$59$]. In this work, we strictly follow the
philosophy of

Tomonaga since it is in formal accord with the Hodge-de Rham theory whose
basics we

describe below.\medskip

We remind to our readers that Hodge-de Rham theory is used thus far in the
Abelian

and non Abelian gauge field theories and, therefore, in the gauge-theoretic
formulations

of gravity.\medskip

\textbf{E.1. Clifford algebra versus second quantization\bigskip }

A quick and very informative introduction to the formalism of second
quantization is

given in books by Tomonaga [$58,59$]. From them, it follows that such a
formalism was

initially designed to treat the processes involving interaction of light
with matter. Since

photons (bosons) are relativistic objects, this requires fermions to be
treated

relativistically as well. That is with help of the Dirac equation. However,
many books

on the second quantization begin with the canonical anticommutation
relations given by 
\begin{equation}
\{a_{i},a_{j}^{+}\}=\delta _{ij},\text{ }\{a_{i},a_{j}\}=0,%
\{a_{i}^{+},a_{j}^{+}\}=0.  \tag{E.1a}
\end{equation}

From these relations the relativistic aspects of the second quantization of
fermions are

not apparent at all! However, following Rosenberg [$60$], we can correct this

deficiency. This is accomplished by introducing the auxiliary operators $%
\hat{e}_{i}=a_{i}-a_{i}^{+},$

$\hat{e}_{i}^{+}=a_{i}+a_{i}^{+}.$ Using these operators along with
Eq.s(E.1a) we obtain at once%
\begin{equation}
\{\hat{e}_{i},\hat{e}_{j}\}=-\{\hat{e}_{i}^{+},\hat{e}_{j}^{+}\}=-2\delta
_{ij},\{\hat{e}_{i},\hat{e}_{j}^{+}\}=0.  \tag{E.2}
\end{equation}

A quick look at the commutators following Eq.(20a) allows us to recognize in
these

anticommutators the already familiar Clifford algebra. This allows us to
define the

Dirac -like operator $d=\dsum\limits_{i}a_{i}^{+}\nabla _{i}$ and, its
adjoint $d^{+}=-\dsum\limits_{i}a_{i}\nabla _{i}$ $\equiv \delta .\footnote{%
In notations of Jost book [$35$] and, \ in view of Eq.s (D.10a), (D.10b),
the same results are written as $d=\varepsilon (\eta _{i})\nabla _{e_{i}}$
and $d^{+}=-i(\eta _{i})\nabla _{e_{i}}.$ Read, please, also the next
subsection.}.$

Here the symbols $d$ and $\delta $ are the raising

and lowering operators of the Hodge-de Rham theory in which the Hodge
Laplacian

$\Delta _{H}$ is acting on differential forms (in the atomic physics these
are the Slater determinants

or their linear combinations) is given by%
\begin{equation}
\Delta _{H}=d\delta +\delta d=\left( d+\delta \right) ^{2}.  \tag{E.3}
\end{equation}%
.

\textbf{E.2. Assortment of the Weitzenbock-Lichnerowicz formulas:
Hartree-Fock }

\ \ \ \ \ \ \ \textbf{versus Hodge-de Rham\bigskip }

The transformation $\hat{e}_{i}=a_{i}-a_{i}^{+},\hat{e}%
_{i}^{+}=a_{i}+a_{i}^{+}$ \ of previous subsection is the simplest case

of Bogoliubov's transformations[$61$], pages 326-336, 527-537. They are
heavily used in

condensed matter and nuclear physics theories. Typically, in such theories
one writes: 
\begin{equation}
\hat{e}_{i}=u_{i}a_{i}-v_{i}a_{i}^{+},\hat{e}%
_{i}^{+}=u_{i}a_{i}+v_{i}a_{i}^{+}  \tag{E.4}
\end{equation}

subject to \ the constraint $u_{i}^{2}$ +$v_{i}^{2}=1.$ In such a case, one
ends up again with the

anticommutator \{$\hat{e}_{i},\hat{e}_{j}^{+}\}=\delta _{ij}$, e.g. see Eq.s
(E.2). In physics this is motivated

by the desire to make transformations, Eq.(E.4), canonical in the sense of

mechanics and quantum mechanics. In mathematics, in the theory of spinors,

the same effect is achieved by selecting either the Clifford or Spinor
bundle.

By selecting between these bundles leads to the assortment of \ the
Weitzenbock-

Lichnerowicz (W-L) formulas. We begin by selecting the Clifford bundle. Such

a choice and the difference between the Clifford and Spinor bundles is nicely

explained in the book by Jost [35], pages 209, 210, 213-218 as well as in the

Appendix D above. This allows us, following [$60]$, to present in this
subsection

the condensed matter-like derivation of the same results. To this purpose,

we select the second quantized \ Hamiltonian \^{H} in the form given on page
528,

Eq.(58.63), of Ref.[$61]$. That is%
\begin{equation}
\text{\^{H}=}\sum_{i}\text{H}_{i}a_{i}^{+}a_{i}-\frac{1}{2}\sum_{i,j,k,l}%
\text{H}_{ijkl}a_{i}^{+}a_{j}^{+}a_{k}a_{l}.  \tag{E.5}
\end{equation}

The obtained results allow us to demonstrate that \^{H} coincides with the
Hodge

Laplacian $\Delta _{H}$. Such a demonstration brings \ the condensed matter
and atomic

physics results in line with those in the Abelian and Non Abelian gauge

field theories.

We begin our demonstration, by using Eq.(D.3). We write: $\theta
(I)=e_{i_{1}}\wedge \cdot \cdot \cdot \wedge e_{i_{k}},$

$I=\{1\leq i_{1}<\cdot \cdot \cdot <i_{k}\leq n\}.$Using Eq.s(D.10a),(D.10b)
it is clear that

$a_{i}^{+}\theta (I)\rightleftarrows \varepsilon (e_{i})\theta (I)$ and $%
a_{i}\theta (I)\rightleftarrows i(e_{i})\theta (I).$ From here, it follows $%
[60]$ that:

a) $\dsum\limits_{i}a_{i}^{+}a_{i}\theta (I)=n\theta (I)$ and

b) if $A^{\ast }$ is an operator inducing an endomorphism of $\theta (I)$%
\begin{equation}
A^{\ast }\theta (I)=\dsum\limits_{j=1}^{k}(-1)^{j}e_{i_{1}}\wedge \cdot
\cdot \cdot \wedge \left( A^{\ast }e_{i_{j}}\right) \wedge \cdot \cdot \cdot
\wedge e_{i_{k}}  \tag{E.6}
\end{equation}

then, provided that $A=A_{ij}$ is skew symmetric, $A^{\ast
}=-\dsum\limits_{ij}A_{ij}a_{j}^{+}a_{j}.$

With these results in our hands and, using the definitions of $d$ and $%
d^{+}, $ the Hodge Laplacian

$\Delta _{H}$ acting on $\theta (I)$ can be presented now as follows: 
\begin{eqnarray}
\Delta _{H}\theta (I) &=&-\dsum\limits_{k,l}(a_{k}^{+}a_{l}\nabla _{k}\nabla
_{l}+a_{l}a_{k}^{+}\nabla _{l}\nabla _{k})\theta (I)  \notag \\
&=&-\dsum\limits_{k,l}(\{a_{k}^{+},a_{l}\}\nabla _{k}\nabla
_{l}-a_{l}a_{k}^{+}(\nabla _{k}\nabla _{l}-\nabla _{l}\nabla _{k}))\theta (I)
\notag \\
&=&\left( -g^{ij}\nabla _{i}\nabla _{j}+\bar{R}\right) \theta (I). 
\TCItag{E.7}
\end{eqnarray}

Here we used Eq.(24). That is, we took into account (see Eq.(E.6)) that

$(\nabla _{k}\nabla _{l}-\nabla _{l}\nabla
_{k})(X)=R(X_{k},X_{l})(X),R(X_{k},X_{l})=-\dsum%
\limits_{i,j}R_{ijkl}a_{i}^{+}a_{j}.$

And, in view of the second line of Eq.(E.7), it is convenient to define $%
\bar{R}=$

$-\dsum\limits_{i,j,k,l}R_{ijkl}a_{i}^{+}a_{j}a_{k}^{+}a_{l}$ . Since
typically $g^{ij}$ is the diagonal matrix and, since

$\dsum\limits_{i}a_{i}^{+}a_{i}\theta (I)=n\theta (I),$by comparing Eq.s
(E.5) and (E.7) and, taking into

account properties a) and b), the identification follows. This provides us
with the

first step toward explaining why Eq.(4) correctly describes the multielectron

atomic system. The task would be completed should $\bar{R}$ in Eq.(E.7) be
replaced by

the scalar curvature $R$. This requires more work leading to the assortment
of the

W-L formulas. In particular, now we are in the position to write down the
1st W-L

formula. Following Ref.[$60]$, the 1st W-L formula is obtained for the
Clifford bundle if

we are interested in using Eq.(E.7) acting on one forms. In such a case,
using Eq.(E.1a)

we obtain $a_{j}a_{k}^{+}=\delta _{jk}-a_{k}^{+}a_{j}$ and then, use it in
the definition of $\bar{R}.$ That is we obtain: 
\begin{equation}
-\dsum\limits_{i,j,k,l}R_{ijkl}a_{i}^{+}a_{j}a_{k}^{+}a_{l}=%
\sum_{il}R_{il}a_{i}^{+}a_{l}-\dsum%
\limits_{i,j,k,l}R_{ijkl}a_{i}^{+}a_{k}^{+}a_{j}a_{l}.  \tag{E.8}
\end{equation}

Here $R_{il}$ \ are the components of the Ricci tensor(read below). The last
term in Eq.(E.8)

naturally produces zero when it is acting on $\theta (I)$ since now it is
the one-form$.$Thus, the

1st W-L formula reads:%
\begin{equation}
\Delta _{H}=\nabla ^{\ast }\nabla +Ric,  \tag{E.9}
\end{equation}

where $\nabla ^{\ast }\nabla =-g^{ij}\nabla _{i}\nabla _{j}$ and, according
to Ref.63, $Ric$=$\sum_{il}R_{il}a_{i}^{+}a_{l}$ represents the Ricci

tensor. This formula was obtained on the Clifford bundle by Jost $[35]$,
page 208, by

slightly different method. The detailed derivation of this result using
formalism of

Clifford algebras is given by Roe[65], page 48. \ In view of developments
presented

in the subsection 4, it is appropriate to describe some fine details of
Roe's derivations

in this subsection. Thus, we begin with the\medskip

\textbf{Definition E.1}. Following [$62$], page 44, we shall call the
combination

$\nabla _{k}\nabla _{l}-\nabla _{l}\nabla _{k}=R(e_{k},e_{l})$ the \textit{%
curvature operator.\medskip }

Using this definition, the following theorem is proven[$62$],\medskip\ pages
46,47:

\textbf{Theorem E.2}. Let $x\rightarrow c(x)$ be a map defined in the
Appendix D 1. Let $M$ be a

Riemannian manifold and $TM$ its tangent bundle. Then, for any $X,Y$ and $Z$ 
$\in TM$

we obtain: 
\begin{equation}
\lbrack R(X,Y),c(Z)]=c(R(X,Y)Z).  \tag{E.10}
\end{equation}

\textbf{Definition E.3}. \ Let $S$ be the Clifford bundle and let $%
K(e_{k},e_{l})$ \ be the \textit{curvature 2-form }

(the same as $R(e_{k},e_{l}))$ \ with values in $End(S).$ Let $e_{i}$ be a
local orthogonal frame on $TM$.

The endomorphism 
\begin{equation}
\mathbf{K}=\dsum\limits_{i<j}c(e_{i})c(e_{j})K(e_{i},e_{j})  \tag{E.11}
\end{equation}

of $S$ is called \textit{Clifford contraction} of $K.$ It is
frame-independent.\medskip

\textbf{Corollary E.4}. Let $\mathcal{D=}d+d^{+}$ be the Dirac (also, the de
Rham operator [$62$], page 51)

operator on the Clifford bundle $S$. Then, the Weitzenbock formula \ is
given by 
\begin{equation}
\mathcal{D}^{2}=\nabla ^{\ast }\nabla +\mathbf{K.}  \tag{E.12}
\end{equation}

Evidently, Eq.(E.9) is a special case of Eq.(E.12).\medskip

\textbf{Definition E.5}. On $TM$ the curvature operator $R$ can be presented
as well via

equation [$62$], page 47,%
\begin{equation}
R(e_{i},e_{j})e_{a}=\dsum\limits_{l}R_{laij}e_{l},  \tag{E.13}
\end{equation}

where $R_{laij}$ is the 4-component Riemann curvature tensor with respect to
the orthogonal

frame made of $e_{i}^{,}$s\medskip

\textbf{Definition E.6. }The $Riemann$ $endomorphism$ $R^{S}of$ $the$ $%
Clifford$ $bundle$ $S$ \ is defined

as 
\begin{equation}
\mathcal{R}^{S}(X,Y)=\frac{1}{4}\dsum%
\limits_{k,l}c(e_{k})c(e_{l})<R(X,Y)e_{k},e_{l}>.  \tag{E.14}
\end{equation}%
\medskip

The $\mathcal{R}^{S}$ is playing central role in \ the W-L-type
calculations. Specifically, by analogy

with Eq.(E.10) it is also possible to arrive at [$62$],page 47,48,%
\begin{equation}
\lbrack \mathcal{R}^{S}(X,Y),c(Z)]=c(R(X,Y)Z).  \tag{E.15}
\end{equation}

The importance of this result can be seen from the Theorem3.16 proven by\
Roe[$62$],

\textbf{Theorem E.7. }The curvature 2-form $\mathbf{K}$ is given by 
\begin{equation}
\mathbf{K}=\mathcal{R}^{S}+\mathcal{F}^{S}  \tag{E.16}
\end{equation}

where $\mathcal{F}^{S}$ is the \textit{twisting curvature} \ of $S$.\medskip

Below, in subsection E.4.,we are going to demonstrate that the twisting
curvature $\mathcal{F}^{S}$

possesses the property%
\begin{equation}
\lbrack \mathcal{F}^{S}(X,Y),c(Z)]=0.  \tag{E.17}
\end{equation}

In the same subsection we shall argue that $\mathcal{F}^{S}\neq 0$ only for
the $spin^{c}$ manifolds and,

therefore, $\mathcal{R}^{S}$ is related to the scalar curvature $R$ while $%
\mathcal{F}^{S}$ is related to the Abelian

curvature $F(A),$ Eq$.(25).\medskip $ Other two \ W-L formulas require [35]
uses of spinor bundles

instead of Clifford bundles. This can be understood if we equivalently
rewrite $\bar{R}=$

$-\dsum\limits_{i,j,k,l}R_{ijkl}a_{i}^{+}a_{j}a_{k}^{+}a_{l}$ as[$60$] 
\begin{eqnarray}
\bar{R} &=&-\frac{1}{16}\dsum\limits_{i,j,k,l}R_{ijkl}(\hat{e}_{i}\hat{e}%
_{j}-\hat{e}_{i}^{+}\hat{e}_{j}^{+})(\hat{e}_{k}\hat{e}_{l}-\hat{e}_{k}^{+}%
\hat{e}_{l}^{+})  \notag \\
&=&\frac{R}{4}+\frac{1}{8}\dsum\limits_{i,j,k,l}R_{ijkl}\hat{e}_{i}\hat{e}%
_{j}\hat{e}_{k}^{+}\hat{e}_{l}^{+}.  \TCItag{E.18}
\end{eqnarray}

The final result is obtained with help of the relation $\hat{e}_{i}\hat{e}%
_{j}-\hat{e}_{i}^{+}\hat{e}_{j}^{+}=-2(a_{i}^{+}a_{j}+a_{i}a_{j}^{+})$ and

taking into account the symmetry of the Ricci tensor: $%
R_{ik}=R_{ki},R_{ik}=R_{ijkj}.$ For details,

see Ref.[$60$],pages 70,71. In Eq.(E.18) $R$ is the scalar curvature. The
combination

$e_{i}e_{j}\bar{e}_{k}\bar{e}_{l}$ is not acting on $\theta (I)$. Instead,
it is acting on $S_{n}$ (defined in Eq.(D.12)),

the 2nd term in Eq.(E.18), when acting on these forms produces zero. Thus,
the

2nd Weitzenbock-Lichnerovicz formula, associated with Eq.(21d), with $%
m^{2}=0,$

reads as follows 
\begin{equation}
\Delta _{H}=\nabla ^{\ast }\nabla +\frac{R}{4}.  \tag{E.19}
\end{equation}

It is given on page 218 of Ref.[$35$], where it was derived differently(read
also Roe $[62]$,

Proposition 3.18., pages 48-49). \ Eq.(4) of the main text is obtained with
help of the

2nd W-L formula while Eq.(25) is obtained \ with help of the 3rd W-L
formula. Actually,

it should be called as Weitzenbock-Lichnerovicz-Schr\"{o}dinger formula,
e.g. Eq.(25), with

$m^{2}=0.$It is presented without proof on page 220 of Ref.[$35$]. This
formula is playing the

central role in the S-W theory discussed in section IV.C. To derive this
formula requires

uses of the concepts of \ spinor and twisted spinor bundle (for $spin^{c}$
manifolds). Even

though Eq.(25) enters \ into the S-W theory [$37$], we need to demonstrate
that the

formalism of atomic physics developed in this work is not only compatible
with the

S-W theory but, in fact, must be looked upon as the special case of this
theory. This

demonstration is presented below. \bigskip

\bigskip \textbf{E.3}. \textbf{BCS superconductivity and the Hodge-de Rham
theory\medskip }

\ \ \ \ \ In section 5, following Witten [$50$], we noticed that the S-W
equations emerge as

generalization of the G-L equations of superconductivity. The microscopic
theory of

superconductivity was initially developed in Ref.[$63]$ by Bardeen,Cooper and

Schriffer (BCS) and, independently, by Bogoliubov [$64$]. Based on
Bogoliubov's results,

Nambu and Jona-Lasino developed a model of elementary particles whose masses
were

generated dynamically. This had been achieved by superimposing the BCS and
the

Dirac equation formalism[$65$]. The detailed derivation of the connection
between the

BCS and the Dirac formalism is presented as Problem P.3.2, on page 80, in
the book [$66$].

Here, we re derive the same results differently for a reason to be explained
in the next

subsection.

\ \ \ \ \ \ \ We begin with Bogoliubov's results following $[67]($Lecture 7.3%
$,$pages 755-772$).$

Also read [$68$]. Start with the Hamiltonian, Eq.(E.5), written in the
reciprocal \textbf{k}-space

as \^{H}=\^{H}$_{0}$+\^{H}$_{V}.$Here \^{H}$_{0}$ stands for one particle
Hamiltonian and, $V$ -for the potential 

energy, that is%
\begin{equation}
\text{\^{H}}_{0}=\dsum\limits_{\mathbf{k}\sigma }\epsilon _{\mathbf{k}}c_{%
\mathbf{k}\sigma }^{\dag }c_{\mathbf{k}\sigma };\text{ \ \^{H}}%
_{V}=\dsum\limits_{\mathbf{kk}^{\prime }}V_{\mathbf{kk}^{\prime }}^{(\mathbf{%
0})}c_{\mathbf{k}}^{\dag }{}_{\uparrow }c_{-\mathbf{k}}^{\dag
}{}_{\downarrow }c_{-\mathbf{k}\downarrow }c_{\mathbf{k}\uparrow }. 
\tag{E.20}
\end{equation}

The spin index $\sigma $ is assumed to have two values: $\uparrow $ and $%
\downarrow .$ In writing \^{H}$_{V}$ only the

potential leading to the spin singlet interactions is present since only
this potential

participates in the superconducting processes. Such written \^{H} is the
Hartree-Fock-type

Hamiltonian[$68$]. The BCS results follow now from the BCS\ assumption that

\begin{equation}
<c_{-\mathbf{k}\downarrow }c_{\mathbf{k}\uparrow }>\neq 0.  \tag{E.21}
\end{equation}

Here \TEXTsymbol{<}$\cdot \cdot \cdot >$ denotes either the quantum
mechanical (zero temperature) or thermal

average. Let, furthermore,%
\begin{equation}
\Delta _{\mathbf{k}}=-\dsum\limits_{\mathbf{k}^{\prime }}V_{\mathbf{kk}%
^{\prime }}^{(\mathbf{0})}<c_{-\mathbf{k}^{\prime }\downarrow }c_{\mathbf{k}%
^{\prime }\uparrow }>.  \tag{E.22}
\end{equation}

Then, \ by applying standard decoupling the effective interaction
Hamiltonian \^{H}$_{V}^{eff}$ \ is

obtained%
\begin{equation}
\text{\^{H}}_{V}^{eff}=-\dsum\limits_{\mathbf{k}}(\Delta _{\mathbf{k}}^{\ast
}c_{-\mathbf{k}\downarrow }c_{\mathbf{k}\uparrow }+\Delta _{\mathbf{k}}c_{%
\mathbf{k}}^{\dag }{}_{\uparrow }c_{-\mathbf{k}}^{\dag }{}_{\downarrow }). 
\tag{E.23}
\end{equation}

Bogoliubov's contribution lies in observations that: a) the anomalous
averages,

Eq. (E.21), lead to nonconservation of the total number of particles (method
of

quasiaverages [$13$],[$14$]), b) the Hamiltonian \^{H}=\^{H}$_{0}+$\^{H}$%
_{V}^{eff}$ is a quadratic form made

of c-operators obeying the anticommutation rules, Eq.(E.1a). Bogoliubov
noticed

that such anomalous (quasi)averages are not typical for superconductivity
only.

They occur in many branches of solid state physics. They play a pivotal role
in this

work as well. Use of method of (quasi)averages is resulting in rigorous 
\textbf{asymptotically}

\textbf{exact solutions} of a variety of quantum many-body problems. To deal
with the

problem b), Bogoliubov proposed to diagonalize the quadratic form \^{H} via
introducing

new $\gamma _{\mathbf{k}\sigma }$ fermionic operators

\begin{eqnarray}
\gamma _{\mathbf{k}\uparrow } &=&u_{\mathbf{k}}c_{\mathbf{k}\uparrow }-v_{%
\mathbf{k}}c_{-\mathbf{k}\downarrow }^{\dag },  \notag \\
\gamma _{\mathbf{k}\downarrow } &=&u_{\mathbf{k}}c_{\mathbf{k}\downarrow
}+v_{\mathbf{k}}c_{-\mathbf{k\uparrow }},  \TCItag{E.24}
\end{eqnarray}

subject to the standard conditions 
\begin{equation}
\{\gamma _{\mathbf{k}\uparrow }^{\dag },\gamma _{\mathbf{k}\uparrow
}\}=1,\{\gamma _{\mathbf{k}\uparrow },\gamma _{-\mathbf{k}\uparrow }\}=0,%
\text{ etc.}  \tag{E.25}
\end{equation}

The diagonalization of \^{H} under such conditions results in the Hamiltonian

\begin{equation}
\text{\^{H}}=\dsum\limits_{\mathbf{k}\sigma }E_{\mathbf{k}}\gamma _{\mathbf{k%
}\sigma }^{\dag }\gamma _{\mathbf{k}\sigma },E_{\mathbf{k}}=\sqrt{\epsilon
_{k}+\left\vert \Delta _{\mathbf{k}}\right\vert ^{2}}.  \tag{E.26}
\end{equation}

With these results presented, we are now in the position to rewrite the
Hodge-de Rham

Laplacian, Eq.(E.8), in the form of Hamiltonian familiar in the solid state
physics. From

the previous subsection we know that $\dsum\limits_{i}a_{i}^{+}a_{i}\theta
(I)=n\theta (I).$ Therefore,

we are dealing with the Hamiltonian \~{H} of the type%
\begin{equation}
\text{\~{H}}=\dsum\limits_{ij}[\tilde{g}^{ij}\nabla _{i}\nabla
_{j}a_{i}^{+}a_{j}-R_{ij}a_{i}^{+}a_{j}],  \tag{E.27}
\end{equation}

where $\tilde{g}^{ij}=\frac{1}{n}g^{ij}.$ This \ is a quadratic form for $a$
operators. It can be diagonalized.

Upon diagonalization the result will look like that in Eq.(E.26), except
that $\tilde{E}_{\mathbf{k}}\neq E_{\mathbf{k}}$. \ 

The question arises: Under what conditions $\tilde{E}_{\mathbf{k}}$ will
look the same as $E_{\mathbf{k}}?$ This can

be achieved based on some auxiliary information from the theory of Dirac
operators.

In physics textbooks the operators $a_{i}^{+},a_{j}$ represent the

creation and annihilation \ operators for electrons while $b_{i}^{+},b_{j}$
are creation and

annihilation operators for positrons (holes). At first look, use of these
operators in the

present context looks permissible but artificial. In the next subsection we
shall

demonstrate how this artificiality disappears for the $spin^{c}$ \
manifolds. For the time

being, we take care of the positron operators in the usual way%
\begin{equation}
\{b_{i},b_{j}^{+}\}=\delta _{ij},\text{ }\{b_{i},b_{j}\}=0,%
\{b_{i}^{+},b_{j}^{+}\}=0.  \tag{E.1b}
\end{equation}

and impose the additional anticommutator relations%
\begin{equation}
\{a_{i},b_{j}\}=\{a_{i},b_{j}^{+}\}=\{a_{i}^{+},b_{j}\}=%
\{a_{i}^{+},b_{j}^{+}\}=0.  \tag{E.1c}
\end{equation}

Next, we assume that the metric $\tilde{g}^{ij}$ in Eq.(E.27) is diagonal
and write instead

of Eq.(E.27), 
\begin{equation}
\text{\~{H}}=\dsum\limits_{i}[\epsilon _{i}a_{i}^{+}a_{i}-\epsilon
_{i}b_{i}^{+}b_{i}]-\dsum%
\limits_{ij}[R_{ij}a_{i}^{+}b_{j}+R_{ij}b_{i}^{+}a_{j}].  \tag{E.28}
\end{equation}

In writing Eq.(E.28) we assumed that only interactions between particles and
holes are

nonzero. In the language of solid state physics, particles are fermions
above the Fermi

surface while the holes are fermions below the Fermi surface.

Taking into account the anticommutation relations, Eq.s(E.1a)-(E.1c), it is
clear that

the quantum system as a whole totally decouples so that it is sufficient to
consider the \ 

diagonalization of the following matrix%
\begin{equation}
M=\left( 
\begin{array}{cc}
\epsilon & -R \\ 
-R & -\epsilon%
\end{array}%
\right)  \tag{E.29}
\end{equation}

resulting in the eigenvalues $E=\pm \sqrt{\epsilon ^{2}+R^{2}}$ coinciding
with those in Eq.(E.26). The

obtained result is in agreement with the result of Nambu and Jona-Lasono[$65$%
]. \ 

Just obtained result demonstrates that Bogoliubov's method of quasiaverages [%
$13$],[$14$]

resulting in obtaining of asymptotically exact diagonalizable model
Hamiltonians can be

translated into the formalism of Hodge-de Rham theory.

It remains to demonstrate that the introduction of the positron operators is
fully

compatible with the 3rd W-L formula, e.g. Eq.(25), paying the central role
in S-W

theory[$37$].\bigskip

\textbf{E.4. \ The BCS superconductivity and the S-W theory. Physics of }$%
spin^{c}$

\ \ \ \ \ \ \ \ \textbf{structures\bigskip }

From the definitions of Clifford and Spinor bundles, Eq.s(D.13), (D.14a),
(D14b), it follows

that, the differences between these bundles is the same as the differences
between the Lie

algebras $so(n)$ and $Spin(n)$ discussed in the Appendix D. $Spin(n)$ is a
double cover of

$so(n)$. Using this fact, it is helpful to restate the content of Eq.(E.6)
as the following\medskip\ 

(Roe $[62]$, Lemma 4.8)\medskip

\textbf{Theorem E.8.}The Lie Algebra $Spin(n)$ can be identified with \ with
the vector subspace

of $Cl(n)$ spanned by the products $e_{i}e_{j},i\neq j.$ The identification
associates the

antisymmetric matrix $A_{ij}$ with the element $\frac{1}{4}%
\dsum\nolimits_{i,j}A_{ij}e_{i}e_{j}\in Cl(n).$ $\medskip $It can be

demonstrated that, $A_{ij}=2(\delta _{i1}\delta _{j2}-\delta _{i2}\delta
_{j1}).\medskip $

Using the Definition E.3., we notice the following. Let $\{e_{k}\}$ be a
local orthonormal

frame for $TM.$ In such a case, the connection and the curvature forms for $%
TM$ \ have their

values in $so(n)$. In particular, the curvature is the $so(n)$-valued
2-form, whose matrix

entries are $(Re_{k},e_{l})$, where $R$ is the Riemann curvature operator,
e.g. see Eq.(E.13).

With help of Theorem E.8. we now obtain, 
\begin{equation}
\mathbf{K}=\frac{1}{4}\dsum\nolimits_{i,j}(Re_{k},e_{l})e_{i}e_{j}, 
\tag{E.30}
\end{equation}

on the Clifford bundle and, 
\begin{equation}
\mathbf{K}=\frac{1}{4}\dsum\nolimits_{i,j}(Re_{k},e_{l})c(e_{i})c(e_{j}), 
\tag{E.31}
\end{equation}

on Spinor bundle. At the same time, use of Eq.s (E.14), (E.16),(E.31),
brings us to

the conclusion that $\mathbf{K}=\mathcal{R}^{S}$ implying that $\mathcal{F}%
^{S}=0.\medskip $

\textbf{Corollary E.9}. The twisting curvature of $Spin(n)$ bundle is
zero.\medskip

If this is so, we need to demonstrate now that only on $Spin^{c}(n)$
manifolds $\mathcal{F}^{S}\neq 0.$

We shall demonstrate this using physics results obtained in previous
subsection.

We begin with the observation that the result, Eq.(E.26), is obtainable if
and only

if in addition to electrons the system also has positrons (holes). That is
the system

is charged, electrically neutral and, hence, the Abelian gauge invariant
initially.

This observation instantly brings up the twisting curvature $\mathcal{F}^{S}$
into play since,

according to Eq.(25), only the twisting curvature is associated with charges.

Thus, the $Spin^{c}(n)$ manifolds should be linked with the charged systems.
Eq.(E.28)

is written for the system of charged fermions. These can be introduced via
the

set of anticommutators, Eq.s(E.1.a)-(E.1.c). Alternatively, instead of using
the

Clifford algebras with the bilinear form $g(x,y)$ having signature
\{1,...,1\}, we can

use the bilinear form with signature \{1,...,1, -1,...,-1\} in which the
number of +1's

is equal to the number of -1's. By analogy with
Eq.s(E.1a),(E.1b),(E.1c),(E.2) we

introduce the additional operators $\hat{E}(e_{i})$ subject to the Clifford
algebra commutation

constraint 
\begin{equation}
\hat{E}(e_{i})\hat{E}(e_{j})+\hat{E}(e_{j})\hat{E}(e_{i})=2\delta _{ij}. 
\tag{E.32}
\end{equation}

Furthermore, we require that 
\begin{eqnarray}
\hat{c}(e_{i})\hat{E}(e_{j}) &=&\hat{E}(e_{j})\hat{c}(e_{i}),\hat{E}%
(e_{j})\equiv E(e_{j}),  \notag \\
\hat{c}(e_{i})\hat{c}(e_{j})+\hat{c}(e_{j})\hat{c}(e_{i}) &=&-2\delta _{ij},%
\hat{c}(e_{i})\equiv c(e_{i}).  \TCItag{E.33}
\end{eqnarray}

Evidently, the operators $\hat{E}(e_{j})$ and $\hat{c}(e_{i})$ are playing \
exactly the same role \ as the

operators $a_{i}$ and $b_{i}$ introduced in Eq.s E(1.b), E(1.c). In complete
analogy with

Eq.(E.31), we define the curvature 2-form $\mathcal{F}^{S}$ as 
\begin{equation}
\mathcal{F}^{S}(e_{i},e_{j})=-\frac{1}{4}\dsum%
\nolimits_{k,l}R_{ijkl}E(e_{k})E(e_{l}).  \tag{E.34}
\end{equation}

At the same time, following $[69],$pages 54-55, we rewrite $\mathcal{R}^{S}$
as 
\begin{equation}
\mathcal{R}^{S}(e_{i},e_{j})=\frac{1}{4}\dsum%
\nolimits_{k,l}R_{ijkl}c(e_{k})c(e_{l}).  \tag{E.35}
\end{equation}

Taking into account Eq.(E.16), we now obtain:%
\begin{eqnarray}
\mathbf{K} &=&\mathcal{R}^{S}(e_{i},e_{j})+\mathcal{F}^{S}(e_{i},e_{j}) 
\notag \\
&=&\frac{1}{4}\dsum\nolimits_{k,l}R_{ijkl}[c(e_{k})c(e_{l})-E(e_{k})E(e_{l})]
\notag \\
&=&-\frac{1}{4}\dsum%
\nolimits_{k,l}R_{ijkl}[c(e_{k})+E(e_{k})][E(e_{l})-c(e_{l})].  \TCItag{E.36}
\end{eqnarray}

A quick look at the Appendix E 2. allows us to write $c(e_{k})%
\rightleftarrows e_{k}=a_{k}-a_{k}^{+},$

$E(e_{k})\rightleftarrows e_{k}=a_{k}+a_{k}^{+}.$ These results are
compatible with the anticommutators

Eq.(E.32),(E.33) since \ Eq.(E.1a) was used for $a_{k}^{,}s.$ Thus, we
obtain:

$[c(e_{k})+E(e_{k})][E(e_{l})-c(e_{l})]=4a_{k}a_{l}^{+}=-4a_{k}^{+}a_{l}.$
Substitution of \ this result

into Eq.(E.36) brings us back to Eq.s (E.8), (E.9) as required. Moreover, the

obtained result is compatible as well with Eq.s(E.28)and (E.29). Now, we are

in the position to finish the description of $Spin^{c}(n)$ manifolds. In
view of the

equivalence $c(e_{k})\rightleftarrows e_{k}=a_{k}-a_{k}^{+}$ and, taking
into account Eq.s(D.10)-(D.12),

just described equivalence can be written also as $c(e_{k})\rightleftarrows
\varepsilon (w)-i(\bar{w})$ while

$E(e_{k})\rightleftarrows \varepsilon (w)+i(\bar{w}).$ The first operator, $%
c(e_{k}),$ is acting on $W$ space

defined \ by Eq.(D.6) while $E(e_{k})$ is operating on $\bar{W}$ space
defined by Eq.(D.8).

Thus, the $Spin^{c}(V)$ $\simeq W\otimes \bar{W}$ in accord with [$70$],
page 512, example 11.1.25.

In view of Eq.(D.13) this is spinor bundle (such that, actually,

$Spin(V)\times S^{1}\longrightarrow Spin^{c}(V))$ in which the associated
bundle is made out of

elements complex conjugate to that in the principal bundle\medskip

\textbf{E.5. Madelung-anomalous superconducting atoms.}

\textbf{\ \ \ \ \ \ \ Superconducting density functional theory (SCDFT).}

\ \ \ \ \ \ \textbf{\ \ Bogoliubov-de Gennes }(\textbf{BdG}) \textbf{%
equations\medskip }

\ \ \ \ SCDFT has been very successful in predicting superconductivity for a
wide variety

of materials, in particular, for study of superconductivity in high pressure
environments

[$5$],[$71$]. According to [$72$], page 9, the BdG equations are completely
analogous to the

SCDFT since they are straightforwardly recoverable from the SCDFT. According
to

[$73$], ch-r 5, the BdG equations are directly connected with (recoverable
from) the

equations of BCS theory of superconductivity. Therefore, it makes sense to
provide

here some details by connecting general results, [$5$],[$73$], with results
of Appendic E.

\ \ \ \ We begin with observation that the SCDFT is built around the
Kohn-Sham (K-S)

density functional theory (DFT). Following [$5$], page 19, we notice that
the key

K-S equation formally coincides with Eq.(3) of this work. We say "formally"
since

Eq.(3) is the Hartree-Fock (H-F) -type of equation in which the potential
contains

the direct and exchange effects [$68$] while in the K-S equation the
potential contains

the material-independent exchange-correlation effects in addition$[5$].
Nevertheless,

using conformal transformations described in section 3.3. of [$3$] it is
possible to

convert Eq.(3) and, hence, the K-S equation, into the equation looking like
our Eq.(4).

But then, according to Appendix E.2, such an equation acquires geometrical

meaning since it can be obtained with help of the 2nd W-L formula.\medskip
This geometrical

meaning is valid for $spin$ manifolds only! To obtain Eq.(25) valid for $%
spin^{c}$ manifolds

requires much more ingenuity as explained in the Appendix E.4. Surprisingly,
only the

ingenuity of purely mathematical results presented in [$62$],[$69$] allows
us to make a

connection with the BdG equations. \ Using Eq.(29), an eigenvalue equation 
\begin{equation}
\left( 
\begin{array}{cc}
\epsilon & -R \\ 
-R & -\epsilon%
\end{array}%
\right) \left( 
\begin{array}{c}
u \\ 
v%
\end{array}%
\right) =E\left( 
\begin{array}{c}
u \\ 
v%
\end{array}%
\right)  \tag{E.37}
\end{equation}

is obtained whose eigenvalues are $E=\pm \sqrt{\epsilon ^{2}+R^{2}}.$ Such
an eigenvalue

equation coincides exactly with the BdG Eq.(5.18) of ch-r 5 of [$73$].
Various

space-time dependent generalizations of Eq.(37) can now be straightforwardly

obtained.\medskip

\textbf{Appendix F}. $\mathit{Spin}$ \textbf{and} $Spin^{c}$\ \textbf{%
structures for Madelung-exceptional atoms}.

\ \ \ \ \ \ \ \ \ \ \ \ \ \ \ \ \ \ \ \ \ \textbf{Simplified treatment}
\medskip

\ \ In Appendix D $spin$ and $spin^{c}$ structures were defined in accord
with their definitions

in mathematical literature. In Appendix E.4. the physical interpretation \
(in terms of

superconductivity concepts) of $spin^{c}$ structures is given for the first
time in mathematical

physics literature. \ However, it is still of interest to check if simple
symmetry rules for

$spin^{c},$defined in section 4.2. are making sense as well. Thus, the
purpose of the

Appendix F lies in checking \ if semi intuitive definitions of section 4.2.
are working.

For this purpose, we believe, use of the following web link

webelements.com/uranium/atoms.html \ \ is helpful.

\ \ \ In this link "uranium" stands for the element uranium. Any other
element is obtained

either by writing its name in the link or by using the l.h.s of \ the web
page for uranium,

where all other elements are listed. For instance, consider the spin
configuration for the

Madelung-regular $_{7}$N ([He] 2s$^{2}2p^{3}),$ that is

\begin{equation}
\fbox{$\uparrow \downarrow $}^{1s^{2}}\fbox{$\uparrow \downarrow $}^{2s^{2}}%
\fbox{$\uparrow $}\fbox{$\uparrow $}\fbox{$\uparrow $}^{2p^{3}}  \tag{F.1}
\end{equation}

From the web link, just given, the levels 1s and 2s are visibly below 2p$.$
Evidently, in the

absence of magnetic field there is spin degeneracy: all "up" spins can be
made "down."

Not surprisingly, this atom is diamagnetic, just like the hydrogen. This is
nematic-type

degeneracy. But, in addition, there is a permutational symmetry. For $_{7}$N
the electrons

at 2s level\ are entangled by the Pauli principle. If one of them is "up"
another must be

"down". Both can be permuted with electrons at 2p level where they are all\ 

indistinguishable. Such permutations are representing the \textbf{additional
symmetry}.

Whenever there is an additional permutational symmetry the atom is
Madelung-regular.

Incidentally, Li, Na, Ka, Rb, Cs are all hydrogen-like and all are
paramagnetic as

noticed already in Section 1.A. Naively, this observation implies that the
"up-down"

symmetry is lost since in case of H it is manifestly present. This fact is
reflected in the

periodic table compiled by Madelung $[17]$ \ who made no comments on this
topic.

The situation, however, is not as simple as it appears in student textbooks,
even those

at the advanced level. The standard theory of Zeeman effect, e.g. read

https://www.damtp.cam.ac.uk/user/tong/aqm/aqmeight.pdf \ 

is telling us that with respect to the static magnetic field all atoms are
both paramagnetic

(this is caused by the interaction term linear in magnetic field, the Pauli
paramagnetic

term) and diamagnetic\ (this is caused by the interaction term quadratic in
the magnetic

field, the Landau diamagnetic term). Experimentally, though, the H atom is
strictly

diamagnetic while the rest of hydrogen-like atoms are strictly paramagnetic.
Apparently,

this fact and, may be, also other factors, e,g. rigorous development of
perturbational

theory for \ superintegrable systems at both classical and quantum levels,
resulted in recent

complete recalculation of the results known in physics literature (other
than [$1,2$],

see also [$43,44$]). The concept of superintegrability is explained in our
work [$3$]. Ref's

[$43,44$] do not include perturbations caused by the relativistic effects.
The sources of

these corrections are described in this paper. At the chemical level of
rigor part of such

calculation was performed in [$45$]. Based on these results, we maintain
that for spins

the "up-down" symmetry is always present but the magnetic properties of
atoms are

the result of all kinds of perturbative effects. Therefore, the overall
paramagnetism

is the cumulative result of these perturbations.\ \ \ \ 

\ \ \ Next, we consider the Madelung-exceptional case, e.g. \ $_{42}$Mo. For
reader's

convenience the list of all Madelung-exceptional atoms is given in

https://en.wikipedia.org/wiki/Aufbau\_principle

Should the Madelung regular rule work, the filling pattern for Mo would be:
[Kr] 4d$^{4}5s^{2}.$

However, the experiment yields: [Kr] 4d$^{5}5s^{1},$that is%
\begin{equation}
\fbox{$\uparrow $}^{5s^{1}}\fbox{$\uparrow $}\fbox{$\uparrow $}\fbox{$%
\uparrow $}\fbox{$\uparrow $}\fbox{$\uparrow $}^{4d^{5}}.  \tag{F.2}
\end{equation}

The levels 4d and 5s are not too distant from each other and, furthermore,
the level 4d is

\textbf{higher }than 5s! There is an obvious "up-down" symmetry for the 4d
level but since

$_{42}$Mo is paramagnetic, the electron on the 5s level should not be mixed
with those on 4d.

The electron at the \textbf{lower} 5s level makes $_{42}$Mo paramagnetic.
This (paramagnetic)

property can be easily seen by removing all 5 of 4d electrons from $_{42}$Mo
resulting in

exactly the same electron configuration as for Rubidium. And Rb is
paramagnetic.

Nevertheless, as for other hydrogenic -like Li, Na, Ka and Cs, the "up-down"

spin symmetry is \textbf{not} lost and the paramagnetism is result of all
kinds of corrections

mentioned above. The case displayed in Eq.(F.1) should be linked with the 
\textit{spin} manifold

because it has permutational symmetry \textbf{additional} to the "up-down"
symmetry. The

case displayed in Eq.(F.2) is linked with the $spin^{c}$ manifold because
the "up-down"

symmetry on 4d level is manifest. Exactly the same analysis is applicable to
the

Madelung-exceptional $_{24}$Cr([Ar] 3d$^{5}4s^{1})$ and to its "twin," Nb.
In the case of Cr, by

stripping it from 5 electrons sitting on the 3d level, we end up with the
configuration

of paramagnetic Potassium, K. Analogously,\ for the Madelung-anomalous Nb

if we strip it from 4 electrons sitting at the 4d level, then we shall bring
the electron

configuration to that of paramagnetic Rubidium, Rb. The Madelung-exceptional
Copper,

Eq.(F.3a), is diamagnetic as well as the Gold since its upper energetic
level 3d is being

occupied by the "Cooper (BCS-type) paired" electrons analogous to that for
the Noble

gases. And all noble gases are diamagnetic as the link

https://periodictable.com/Properties/A/MagneticType.html \ \ \ demonstrates.

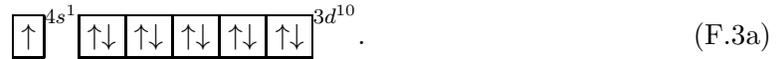
\begin{equation}
\fbox{$\uparrow $}^{4s^{1}}\fbox{$\uparrow \downarrow $}\fbox{$\uparrow
\downarrow $}\fbox{$\uparrow \downarrow $}\fbox{$\uparrow \downarrow $}\fbox{%
$\uparrow \downarrow $}^{3d^{10}}.  \tag{F.3a}
\end{equation}

Here, the 4s level is lover than 3d. The pattern for the Gold is similar%
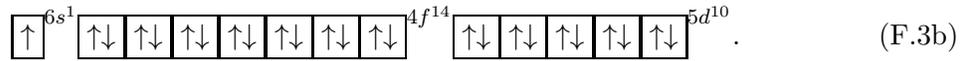
\begin{equation}
\fbox{$\uparrow $}^{6s^{1}}\fbox{$\uparrow \downarrow $}\fbox{$\uparrow
\downarrow $}\fbox{$\uparrow \downarrow $}\fbox{$\uparrow \downarrow $}\fbox{%
$\uparrow \downarrow $}\fbox{$\uparrow \downarrow $}\fbox{$\uparrow
\downarrow $}^{4f^{14}}\fbox{$\uparrow \downarrow $}\fbox{$\uparrow
\downarrow $}\fbox{$\uparrow \downarrow $}\fbox{$\uparrow \downarrow $}\fbox{%
$\uparrow \downarrow $}^{5d^{10}}.  \tag{F.3b}
\end{equation}

The level 5d has the highest \ energy while the 6s - the lowest. The
analysis of the

Madelung-exceptional Platinum, Rodium and Ruthenium proceeds analogously.

It is somewhat trickier though. Consider, for instance, the Ruthenium%
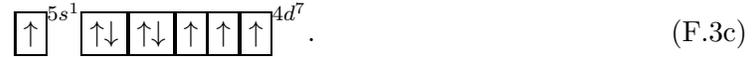
\begin{equation}
\fbox{$\uparrow $}^{5s^{1}}\fbox{$\uparrow \downarrow $}\fbox{$\uparrow
\downarrow $}\fbox{$\uparrow $}\fbox{$\uparrow $}\fbox{$\uparrow $}^{4d^{7}}.
\tag{F.3c}
\end{equation}

Its analysis proceeds in a very much the same way as that for Mo since,
unlike the N,

whose configuration is displayed in Eq.(F.1), all energies of electrons at
4d level are

the same. That is the "Cooper paired" and unpaired electrons are sitting at
the same 4d

energy level which is visibly higher that 5s level. The treatment of the
remaining

Madelung-exceptional Palladium remains is puzzling. Its electronic
configuration

apparently implies that it should be diamagnetic, like Gold, but it is
paramagnetic!

Nevertheless, from the point of view of the "up-down" symmetry it is surely
the

Madelung-exceptional.

\ \ \ Next, we want to comment on Madelung-exceptional lantanides (La and
Ce) and

actinides (Ac and Th). For $_{57}$La the standard Madelung rule prescribes
the configuration

[Xe]$4f^{1}5d^{0}6s^{2}$ while the experiment provides [Xe]$%
4f^{1}5d^{1}6s^{2},$ that is,%
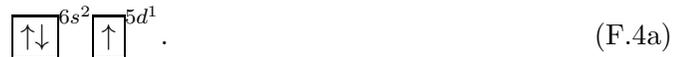
\begin{equation}
\fbox{$\uparrow \downarrow $}^{6s^{2}}\fbox{$\uparrow $}^{5d^{1}}. 
\tag{F.4a}
\end{equation}

As before, the level 6s is \textbf{lover} than 4f and this level is lover
than 5d. $_{57}$La is behaving

the same way as the rest of hydrogen-like atoms and, therefore, it is
paramagnetic as

the rest of them. The electronic configuration of $\ _{58}$Ce is: [Xe]$%
4f^{0}5d^{1}6s^{2}.$ It is

paramagnetic, as expected, so that its hydrides should have properties very
much

analogous to those of $_{57}$La. Next, for \ $_{89}$Ac we have the situation
mirroring that of $_{57}$La,

except that the orbital energy levels are higher: [Rn]$5f^{0}6d^{1}7s^{2}.$
Finally$,$for the Thorium,

$_{90}$Th, we have: [Rn]$4f^{0}6d^{2}7s^{2},$ that is 
\begin{equation}
\fbox{$\uparrow \downarrow $}^{7s^{2}}\fbox{$\uparrow $}\fbox{$\uparrow $}%
^{6d^{2}}  \tag{F.4b}
\end{equation}

with 7s energy noticeably lower than 6d so that, again, we have spin$^{c}$
manifold. We have

$_{90}$Th being paramagnetic analogously to all hydrogen-like atoms. It is
acting like\ Madelung-

exceptional Rhodium, that is 
\begin{equation}
\fbox{$\uparrow $}^{5s^{1}}\fbox{$\uparrow \downarrow $}\fbox{$\uparrow
\downarrow $}\fbox{$\uparrow \downarrow $}\fbox{$\uparrow $}\fbox{$\uparrow $%
}^{4d^{8}}.  \tag{F.5}
\end{equation}

All lantanides and actinides are paramagnetic, including Gadolinium,

http://mriquestions.com/why-gadolinium.html,

even though another table

https://periodictable.com/Properties/A/MagneticType.html

states that Gd is ferromagnetic.

Finally, let us take a look at the Sulphur, S, whose nonmetallic hydride

demonstrated the highest to date T$_{c}$ under high pressures. For $_{16}$S
we have 
\begin{equation}
\fbox{$\uparrow \downarrow $}^{3s^{2}}\fbox{$\uparrow \downarrow $}\fbox{$%
\uparrow $}\fbox{$\uparrow $}^{3p^{4}}  \tag{F.5}
\end{equation}

Eq.(F.5) clearly exhibits the "up-down" and permutational symmetry thus
making $_{16}$S

Madelung-regular and diamagnetic. Two unpaired electrons make $_{16}$S \ to
act as if it is

Madelung-exceptional $_{90}$Th, thus assuring its high temperature
superconducting

capabilities. These were indeed observed. The difference in the atomic masses

positively affected the observed T$_{c}^{^{\prime }}s$.\medskip

\bigskip

\bigskip

\bigskip

\bigskip

\textbf{References}

\bigskip

$[1]$ \ \ A.Rau, $\func{Re}$p.$\Pr $ogr.Phys. \textbf{53,} 181 (1990).

$[2]$ \ \ H. Friedrich, D.Wintgen, Phys.Reports \textbf{183, }37 (1989).

$[3]$ \ \ A.Kholodenko, L. Kauffman, \ "How the modified Bertrand theorem

\ \ \ \ \ \ \ \ explains\ regularities of the periodic table I. From
conformal

\ \ \ \ \ \ \ \ invariance to Hopf mapping," arXiv:1906.05278.

[$4$] \ \ E.Wigner, H.Huntington, J.Chem.Phys.\textbf{140}, 764 (1935).

$[5]$ \ \ J.Flores-Livas, L. Boeri, A.Sanna, G.Profeta, R. Arita, M. Eremets,

\ \ \ \ \ \ \ Phys.Reports \textbf{856,}1 (2020).

[$6$] \ \ J. Gilman, Phys.Rev.Lett.\textbf{26}, 546 (1971).

[$7$] \ \ S.Setayandeh,C.Webb, E. MacA Gray, \ Progr.Solid St.Chem. \textbf{%
60},100265 (2020).

[$8$] \ \ J.Gordon, H.Montgomery, R.Noer,G.Pickett, R.Tobon.Phys.Rev. 
\textbf{152},432 (1966).

[$9$] \ \ C.Satterthwite, L.Toepke, Phys.Rev.Lett.\textbf{25},741 (1970).

[$10$] \ A.Z\"{u}ttel, materialstoday \textbf{6}, 24 (2003).

[$11$] \ B.Adams, A.Chen, materialstoday \textbf{14}, 282 (2011).

[$12$] \ S.Lidquist, N.March, \textit{Theory of the Inhomogenous Electron Gas%
}

\ \ \ \ \ \ \ (Springer Science+Business Media, New York,1983).

[$13$] \ N.Bogoliubov, \textit{Lectures on Quantum Statistics}, Vol.2

\ \ \ \ \ \ \ (Gordon and Breach Sci.Publishers, New York, 1970).

[14] \ N.Bogoliubov, N.Bogoliubov Jr., \textit{Introduction to Quantum
Statistical Mechanics}

\ \ \ \ \ \ \ (World Scientific, Singapore, 2010).

$[15]$ \ P.Thyssen, A. Ceulemans, \textit{Shattered Symmetry}

\ \ \ \ \ \ \ (Oxford University Press, Oxford, 2017).

$[16]$ \ H.Bethe, R.Jackiw, \textit{Intermediate Quantum Mechanics}

\ \ \ \ \ \ \ (CRC Press, New York, 2018).

$[17]$ \ E.Madelung, $\mathit{Die}$ $\mathit{Mathematischen}$ $\mathit{%
Hilfsmittel}$ $\mathit{des}$ $\mathit{Physikers}$

\ \ \ \ \ \ (Springer-Verlag, Berlin,1936).

$[18]$ \ M. Englefield, \textit{Group Theory and Coulomb Problem\ }

\ \ \ \ \ \ \ (Wiley Interscience, New York,1972).

$[19]$ \ Y.Demkov, V.Ostrovsky, Sov.Phys. JETP \textbf{13,}1083 (1971).

$[20]$ \ H.Goldstein, C. Poole, J. Safko, \textit{Classical Mechanics}

\ \ \ \ \ \ (Pearson Education Ltd, London, 2014).\textit{\ }

$[21]$ \ M.Gutzviller,\textit{\ Chaos in Classical and Quantum Mechanics}

\ \ \ \ \ \ (Springer-Verlag, Berlin, 1990).

$[22]$ \ R. Luneburg, \textit{Mathematical Theory of Optics}

\ \ \ \ \ \ (U.of California Press, Los Angeles,CA, 1966).

$[23]$ \ V. Perlick, Class.Quantum Grav\textit{.} \textbf{9, }1009 (1992).

$[24]$ \ K. Dyall, K. Faegri, \textit{Introduction to Relativistic Quantum}

\ \ \ \ \ \ \ \textit{Chemistry} (Oxford U.Press, Oxford, 2007).

$[25]$\ \ S.Singer, \textit{Linearity, Symmetry and Prediction in the
Hydrogen Atom}

\ \ \ \ \ \ \ (Springer-Verlag,Berlin, 2005).

$[26]$ \ P. Martin, R.Glauber, Physical Review \textbf{109,}1307 (1968).

$[27]$ \ L. Biedenharn, Found.Phys. \textbf{13},13 (1983).

$[28]$ \ P. Dirac, \textit{Principles of Quantum Mechanics}

\ \ \ \ \ \ \ \ (Clarendon Press, Oxford, 1958).

$[29]$\ \ P. Collas, Am.J. of Phys\textit{.}\textbf{38, }253 (1978).

$[30]$\ \ E. Schr\"{o}dinger, Sitzunsber.Preuss.Acad.Wiss. Phys.Math\textit{%
. }\textbf{K1,}105\textbf{\ }(1931).

$[31]$\ \ F.Hehl, P.Heyede, G. Kerlick, Rev.Mod.Phys. \textbf{48, }393%
\textbf{\ }(1976).

$[32]$ \ B. Kay, Gen. Relativity and Gravitation \textbf{52, }1 (2020).

$[33]$ \ J.Moore, \textit{Lectures on Seiberg-Witten invariants}

\ \ \ \ \ \ \ (Springer-Verlag, Berlin,1996).

[$34$] \ A. Kholodenko, L. Kauffman, \ Ann.Phys. \textbf{390,}1 (2018).

$[35]$ \ J. Jost, \textit{Riemannian Geometry and Geometric Analysis}

\ \ \ \ \ \ \ (Springer-Verlag, Berlin, 2005).

$[36]$ \ S. Kuru, J. Negro, O. Ragnisco, Phys.Lett.\textit{\ }A \textbf{381, 
}3355\textbf{\ (}2017\textbf{)}.

[$37$] \ S.Donaldson, BAMS \textbf{33, }45\textbf{\ (}1996\textbf{).}

[$38$] \ C. Itzykson, J-B. Zuber, \textit{Quantum Field Theory}

\ \ \ \ \ \ \ (McGraw-Hill, New York, 1980).

[$39$] \ M.Wong, H-Y.Yeh, Phys.Rev. D\textit{\ }\textbf{25, }3396 (1982).

[$40$] \ G.Naber, \textit{Topology, Geometry and Gauge Fields: Interactions}

\ \ \ \ \ \ \ (Springer Science-Business Media, Berlin, 2011).

$[41]$\ \ M. Nakahara, \textit{Geometry,Topology and Physics}

\ \ \ \ \ \ \ (Adam Hilger, New York,1990).

$[42]$ \ L.Nicolaescu, \textit{Notes on the Atiyah-Singer Index Theorem}

\ \ \ \ \ \ \ (https://www3.nd.edu/\symbol{126}lnicolae/ind-thm.pdf \ 2013).

[$43$] \ F.Fasso, D.Fontanari, D.Sadovskii, Math.Phys.Anal.Geom.\textbf{18, }%
30 (2015).

[$44$] \ K.Etstathiou, D.Sadovski, Rev.Mod. Phys. \textbf{82, }2099 (2010).

[$45$] \ H.Tatewaki, S.Yamamoto,Y. Hatano, Comp.\&Theoret.Chemistry \textbf{%
1125}, 49 (2018).

$[46]$ \ M.Blume, R.Watson, Proc.Roy.Soc.(London) A \textbf{270, }127 (1962).

$[47]$ \ M.Blume, R.Watson, A.Freeman, Phys.Rev.\textbf{134, }320 (1964).

$[48]$ \ L.Schiff, \ \textit{Quantum Mechanics} (McGraw-Hill, New York,1968).

$[49]$ \ G.Naber, \textit{Geometry, Integrability and Quantization},
pp181-199,

\ \ \ \ \ \ \ (Coral Press Sci.Publ. Sofia, 2000).

$[50]$ \ E.Witten, BAMS \textbf{44, }361 (2007).

$[51]$ \ C.Taubes,\textit{\ Seiberg-Witten and Gromov Invariants for
Symplectic\ }

\ \ \ \ \ \ \ \textit{4-Manifolds} (International Press, Somerville, MA,
2000).

$[52]$ \ A. Kholodenko, \textit{Applications of Contact Geometry and Topology%
}

\ \ \ \ \ \ \ \textit{in Physics\ }(World Scientific, Singapore, 2013).

$[53]$ \ \ C.Adam, B.Muratori, C.Nash, JMP \textbf{41, }5875\textbf{\ }%
(2000).

$[54]$ \ A.Sergeev, \textit{Vortices and Seiberg-Witten Equations},

\ \ \ \ \ \ \ Lecture Notes (Nagoya University, 2009).

[$55$] \ J.Morgan, \textit{The Seiberg-Witten Equations and Applications}

\textit{\ \ \ \ \ \ to the Topology of Smooth Four Manifolds}

\textit{\ \ \ \ \ \ }(Princeton U.Press, Princeton, 1996).

$[56]$ \ E.Witten, Math.Research Lett.\textbf{1, }769 (1994).

$[57]$ \ S.Tomonaga,\textit{\ Quantum Mechanics},Vol.II,

\ \ \ \ \ \ \ (John Wiley \& Sons, Inc. New York, 1966).

$[58]$ \ S.Tomonaga, \textit{The Story of Spin,}

\ \ \ \ \ \ \ (U.of Chicago Press, 1997) .

$[59]$ \ F.Harris, H.Monkhorst, D.Freeman, \textit{Algebraic and Diagrammatic%
}

\ \ \ \ \ \ \ \textit{Methods in Many-Fermion Theory }(Oxford U.Press,
Oxford, 1992).

$[60]$ \ S.Rosenberg, \textit{The Laplacian on a Riemannian Manifold}

\ \ \ \ \ \ \ (Cambridge U.Press, Cambridge, 1997).

$[61]$ \ A.Feter, J.Walecka, \textit{Quantum Theory of Many-Particle Systems}

\ \ \ \ \ \ \ (McGraw Hill, New York, 1971).

$[62]$ \ J.Roe, \textit{Elliptic Operators,Topology and Asymptotic Methods}

\ \ \ \ \ \ (Chapman\&Hall/CRC, New York, 2001).

$[63]$ \ J.Bardeen, L.Cooper, J.Schrieffer, Phys.Rev.\textbf{108, }1175
(1957).

$[64]$ \ N. Bogoliubov, V.Tolmachev, D.Shirkov, \textit{A New Methodd in\ }

\textit{\ \ \ \ \ theTheory of Superconductivity }(Consultants Bureau, Inc.,
Washington, 1959).

$[65]$ \ Y.Nambu, G.Jona-Lasino, Phys.Rev.\textbf{122, }345\textbf{\ }(1961).

$[66]$ \ J-P. Blaizot, G.Ripka, \textit{Quantum Theory of Finite Systems}

\ \ \ \ \ \ (MIT Press, Cambridge, MA 1986).

$[67]$ \ C.Henley, \textit{Condensed Matter Physics} (unpublished book),

\ \ \ \ \ \ Lecture 7.3. BCS Theory, Cornell University, 2009.

$[68]$ \ H.Simon, \textit{Lecture Notes for Quantum Matter},

\ \ \ \ \ \ \ (Oxford University,\ Oxford, 2019).

$[69]$ \ X.Dai, \textit{Lectures on Dirac Operators and Index Theory}

\ \ \ \ \ \ \ (UCSB, January 7, 2015).

[$70$] \ L.Nicolaescu, \textit{Lectures on the Geometry of Manifolds}, 2nd
edition,

\ \ \ \ \ \ (World Scientific, Singapore, 2007).

[$71]$ \ L.Oliviera, E.Gross, W.Kohn, PRL\textbf{\ 60}, 2340 (1988).

[$72$] \ J.Schmidt, C.Benavides-Riveros, M.Marques, Phys. Rev. \textbf{B 99}%
, 224502 (2019).

[$73$] \ P.De Gennes, \textit{Superconductivity of Metals and Alloys}

\ \ \ \ \ \ (CRC Press, Taylor \& Francis group, Boca Raton, FL 2018).

[$74$] \ P.Anderson, J.Phys.Chem.Solids \textbf{11}, 28 ( 1959).

[$75$] \ J.Non Delft, D. Ralph, Phys.Reports \textbf{345}, 61 (2001).

[$76$] \ A.Kholodenko, J.Geom.Phys.\textbf{59}, 600 (2009).

[$77$] \ A.Kholodenko, Int.J.Geom.Methods in Mod.Phys. \textbf{8}, 1355 (
2011).

[$78$] L.Carrier, Ch.-E Fectau, P.Johnson, Int.J.Quantum Chem. \textbf{120},
26255 (2020).

[$79$] \ D.Freed, K.Uhlenbeck, Instantons and Four Manifolds

\ \ \ \ \ \ (Springer-Verlag, Berlin, 1984).

[$80$] \ R.Richardson, N.Sherman, Nucl.Phys.\textbf{62}, 221 (1964).

[$81$] \ R.Richardson, JMP \textbf{9}, 1327 (1968).

[$82$] \ L.Cooper, Phys.Rev.\textbf{104}, 1180 (1956).

[$83$] \ M.Sambataro, N.Sandulescu, Ann.Phys. \textbf{413},168061 (2020).

[$84$] \ V.Kresin,Yu.Ovchinnikov, Ann.Phys.\textbf{\ 417}, 168141 (2020).

[$85$] \ N.Kuzmenko, Physica C \textbf{576}, 1353709 (2020).

[$86$] P.Edwards, M.Khojasteh, A.Halder, V.Kresin, J.of Superconductivity and

\ \ \ \ \ \ \ Novel Magnetism doi.org./10.1007/s10948-021-0602-y

[$87$] \ P.Jena, C.Satterthwaite, \textit{Electronic Structure and
Properties of Hydrogen in Metals}

\ \ \ \ \ \ (Plenum Press, New York, 1983).

[$88$] \ D.Broom, \textit{Hydrogen Storage Materials }

\ \ \ \ \ \ (Springer-Verlag, London Ltd., 2011).

[$89$] \ D.Dean, M.Hjorth-Jensen, Rev.Mod.Phys. \textbf{75}, 607 (2003).

[$90$] \ N.Watari, S.Ohnishi, Y.Ishii, J.Phys.Cond.Matter \textbf{12}, 6799
(2000).

\bigskip

\bigskip

\bigskip

\bigskip

\ \ \ \ \ \ \ \ \ \ \ \ \ \ \ \ \ \ 

\ \ \ \ \ \ \ \ 

\bigskip

\bigskip

\bigskip

\bigskip

\bigskip

\bigskip

\bigskip

\bigskip

\bigskip

\bigskip

\bigskip

\ \ 

\end{document}